\RequirePackage{fix-cm}
\documentclass[twocolumn]{svjour3}          % twocolumn
\smartqed  % flush right qed marks, e.g. at end of proof

\usepackage{mwe}
\usepackage{multirow,makecell,array}
\usepackage{subfig}
\usepackage{graphicx}
\usepackage{caption}
\usepackage{amsmath,amsfonts,amssymb}
\usepackage{float,natbib}
\usepackage{color}
\usepackage{float}
\usepackage{cuted}

\newcommand{\OO}{\mathrm{\bf O}}
\newcommand{\AAA}{\mathrm{\bf A}}
\newcommand{\PPP}{\mathrm{\bf P}}
\newcommand{\ee}{\mathrm{\bf e}}
\newcommand{\uu}{\mathrm{\bf u}}
\newcommand{\BB}{\mathrm{\bf B}}
\newcommand{\CC}{\mathrm{\bf C}}

\begin{document} 

\title{Size-dependency and lattice-discreetness effect on fracture toughness in 2D crystals under antiplanar loading} 


\author{Thuy Nguyen         \and
        Daniel Bonamy %etc.
}


\institute{T. Nguyen \at
              Léonard de Vinci Pôle Universitaire, Research Center, 92 916 Paris La Défense, France \\
              Tel.:\\
              Fax: \\
              \email{thuy.nguyen@devinci.fr}           %  \\
           \and
           T. Nguyen \and D. Bonamy \at
              Service de Physique de l'Etat Condensée, CEA, CNRS, Université Paris-Saclay, CEA Saclay 91191 Gif-sur-Yvette Cedex, France\\
              Tel.:\\
              Fax: \\
              \email{daniel.bonamy@cea.fr}    
}

\date{Received: date / Accepted: date}

\maketitle

\begin{abstract}
Fracture toughness is the material property characterizing resistance to failure. Predicting its value from the solid structure at the atomistic scale remains elusive, even in the simplest situations of brittle fracture. We report here numerical simulations of crack propagation in two-dimensional fuse networks of different periodic geometries, which are electrical analogs of bidimensional brittle crystals under antiplanar loading. Fracture energy is determined from Griffith's analysis of energy balance during crack propagation, and fracture toughness is determined from fits of the displacement fields with Williams' asymptotic solutions. Significant size dependencies are evidenced in small lattices, with fracture energy and fracture toughness both converging algebraically with system size toward well-defined material-constant values in the limit of infinite system size. The convergence speed depends on the loading conditions and is faster when the symmetry of the considered lattice increases. The material constants at infinity obey Irwin's relation and properly define the material resistance to failure. Their values are approached up to $\sim 15\%$ percent using the recent analytical method proposed in \cite{Nguyen19_prl}. Nevertheless, the deviation remains finite and does not vanish when the system size goes to infinity. We finally show that this deviation is a consequence of the lattice discreetness and decreases when the super-singular terms of Williams' solutions (absent in a continuum medium but present here due to lattice discreetness) are taken into account.
\end{abstract}

\section{Introduction}
\label{intro}

Crack growth drives brittle material failure. Linear Elastic Fracture Mechanics (LEFM) provides the relevant framework to describe the problem. It is based on the fact that externally applied stresses are concentrated by a crack, to values that become mathematically singular at the crack tip. The near-tip field, then, is fully characterized by a single parameter, coined stress intensity factor, $K$ \citep{Lawn93_book,Bonamy17_crp}. Crack growth initiates when the energy release rate, $G$, i.e. the mechanical energy released as the crack propagates over a unit length, becomes larger than the fracture energy, $\Gamma$, i.e. the energy dissipated to expose a new unit area of fracture surfaces \citep{Griffith20_ptrs}. The former connects to $K$ via the relation of \cite{Irwin57_jam}: $G = K^2/E$ where $E$ is Young's modulus. The latter connects to fracture toughness, $K_c = \sqrt{\Gamma E}$, which characterizes the material resistance-to-failure.

LEFM theory allows the determination of $G$ and $K$, at least numerically, for any geometry. On the other hand, efforts to relate $\Gamma$ and $K_c$ to the atomistic parameters have been only partially successful; they are considered as material constants, to be determined experimentally. However, it should {\em a priori} be possible to infer their value {\em ab-initio}, at least for perfectly brittle solids in which the crack advance proceeds through successive bond breaking, without involving additional elements of dissipation (dislocation, crazing, etc). In his seminal work, \cite{Griffith20_ptrs} proposed to identify $\Gamma$ with the free surface energy per unit area, $\gamma_S$, which is the bond energy times the number of bonds per unit surface: $\Gamma = 2 \gamma_S$. However, fracture energy measured in many different materials (even the most brittle ones) is always found significantly higher \citep{Perez00_prl,Zhu04_prl,Buehler07_prl,Kermode15_prl}. 

Several reasons have been proposed to explain this discrepancy. \cite{Thomson71_jap} have invoked the lattice trapping effect due to the solid discreetness at the atomic scale, with additional finite energy barriers to overcome to break the successive bonds \citep{Hsieh73_jap,Marder95_jmps,Bernstein03_prl, Santucci04_prl}. \cite{Slepyan81_spd, Kulakhmetova84_ms, Slepyan10_ap} have invoked the existence of high-frequency phonons to take into account in the energy budget to determine $\Gamma$. Finally, \cite{Nguyen19_prl} proposed that $K_c$ originates in the way the near-tip mathematical singularity of the stress field is positioned in the discrete atomic lattice. Still, it remains difficult to assess the mechanisms above in fracture experiments, due to the lack of control over the material parameters. 

Here, we reported a series of numerical simulations performed on minimal fracturing systems: two-dimensional (2D) fuse networks of modulated sizes and geometry. These systems are electrical analogs of 2D perfectly brittle crystals subjected to antiplanar loading. The numerical scheme is detailed in Section 2. The analysis of the voltage field (analog of the out-of-plane displacement) and its fitting with Williams' asymptotic expansion allow the accurate determination of fracture toughness and relative contribution of higher-order sub-singular terms (Section 3). The examination of the energy balance as the crack advances enables the determination of fracture energy (Section 4). The dependency of these fracture parameters with system size and geometry is characterized in Section 5: Fracture toughness and fracture energy both converge toward material-constant values as system size goes infinite. The convergence speed depends on loading conditions and is faster when the symmetry of the considered lattice increases. Section 6 compares these continuum-level scale material constant values to analytical predictions obtained using the procedure developed by \cite{Nguyen19_prl}, which consists in using Williams's expansion supplemented with an extra super-singular term to accurately position the crack tip in the discrete lattice, and subsequently prescribing stress intensity factor so that the force applying onto the next bond to break is at the breaking threshold. Analytical predictions are compatible with numerical ones within $\sim 15$ percent. The remaining deviation is demonstrated to be a consequence of the lattice discreetness and can be decreased by considering the super-singular terms of Williams' solutions.

\section{Numerical method}
\label{sec:numeticalmethod}

\begin{figure*}[tp]
  \includegraphics[width=\textwidth]{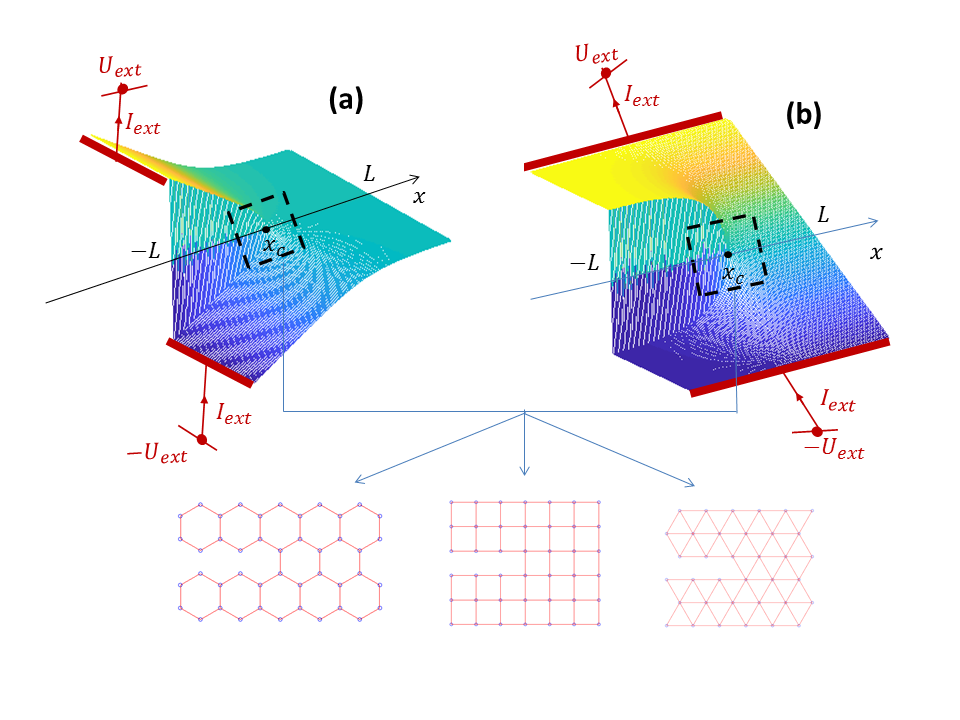}
\caption {Sketch and notations of the fuse lattices simulated here in so-called CT  (panel {\bf a}) and TS (panel {\bf b}) configurations. External voltage difference $2 U_{ext}$ is imposed between the upper and lower part of the left-handed edge in the first case and between the horizontal top edge and the bottom one in the second case. In response, a total current $I_{ext}$ flows through the network. System size is $2L\times L$ in the $x$ and $y$ directions so that the $x$ axis is parallel to the crack. For each loading configuration, three different geometries are examined: Honeycomb, square, and triangular.}
\label{fig:1}       % Give a unique label
\end{figure*}

\begin{table}[htp]
    \centering
\begin{tabular}{|l|c|c|c|}
  \hline
   & Honeycomb & Triangular & Square  \\
  \hline
   $E$ & $2/\sqrt{3}$    &  $2\sqrt{3}$   & 2\\
  \hline
  $\gamma_S$ & $1\sqrt{3}$ & 2 & 1 \\
  \hline
\end{tabular}
\caption{Young's modulus, $E$, and free surface energy, $\gamma_S$, in the three different lattice geometries investigated here. Both are expressed in reduced units: $g$ unit for Young's modulus and $\gamma / \ell$ units for free surface energy. $g$, $\gamma$, and $\ell$ denote the fuse conductance, the binding energy, and the fuse length, respectively}
\label{tab:Egamma}
\end{table}

The simulated system is depicted in Fig. \ref{fig:1}. In the following,  We employ fuse lattices to model crack propagation in bidimensional materials subjected to antiplanar loading \citep{Arcangelis89,Hansen91_prl,Zapperi97_nature,Zapperi05_pre}. Indeed there is a formal analogy between scalar elasticity and electrical problems, so that the local voltage, $u(x,y)$, corresponds to the out-of-plane displacement, $\vec{u}=u(x,y)\vec{e}_z$, the current flowing through each fuse, $i$, corresponds to the force acting on the associated bond, the fuse conductance, $g$, corresponds to the bond stiffness, and the fuse breakdown, $i_c$, corresponds to the bond strength \citep{Arcangelis89,Hansen91_prl,Zapperi97_nature,Zapperi05_pre}. In the following, $g = 1$, and all fuses have a unit length, $\ell=1$. Moreover, $i_c$ is set $i_c = \sqrt{2}$, so that the corresponding binding energy $\gamma = (1/2)i_c^2/g$ is equal to $1$. 

Three different lattice geometries are considered: honeycomb, square, and triangular.
 All slabs employed for simulations have a dimension $2L \times L$ in the $x$ and $y$ directions respectively. A horizontal (along $x$) straight crack of initial length $4L/5$ is introduced from the left-handed side in the middle of the strip, by withdrawing the relevant fuses (Fig. \ref{fig:1}). Two different loadings are imposed: 
\begin{itemize}
    \item[(i)] Compact tension (CT) loading, generated by imposing a voltage difference $2U_{ext}$ between the upper (above the crack) and lower (below the crack) portions of the left edge (Fig. \ref{fig:1}(a)).
    \item[(ii)] Thin strip (TS) loading, generated by imposing a voltage difference $2U_{ext}$ between the top and bottom edges of the slab (Fig. \ref{fig:1}(b)).
\end{itemize}

The voltage at each node is then obtained by solving Kirchhoff's and Ohm's laws. For both loading schemes, $U_{ext}$ is increased till the moment $t$ where the current flowing in the unbroken fuse at the crack tip is equal to $i_c$; this threshold value $U_{ext}(t)$ corresponds to the crack propagation onset. The unbroken fuse at the crack tip is then burnt and the crack advances over one step. Kirchhoff's and Ohm's law is then used again to compute voltage at each node, $U_{ext}$ is adjusted again to the next burning point, and so on. This process is repeated till the crack has propagated over an additional distance of $2L/5$. In the following, the x-coordinate $x_c$ refers to the center of mass of the fuse at the crack tip about to burn. 

In the simulations presented here, the system size varied from $L=10$ up to $L=200$. In the following, all quantities are expressed in reduced units: $\ell$ unit for length, $\sqrt{g \gamma}$ unit for current, $\sqrt{\gamma/g}$ unit for voltage, $g$ unit for the elastic modulus of the 2D lattices, $\gamma/\ell$ unit for surface or fracture energy, and $\sqrt{g\gamma / \ell}$ unit for stress intensity factor or fracture toughness. Note finally that, in the three different lattice geometries considered here, Young's modulus and free surface energy are known \citep{Nguyen19_prl}; their value is reported in Tab. \ref{tab:Egamma}.  

\section{Displacement field analysis}
\label{sec:displacementField}

\begin{figure*}[htp]
 \includegraphics[width=0.99\textwidth]{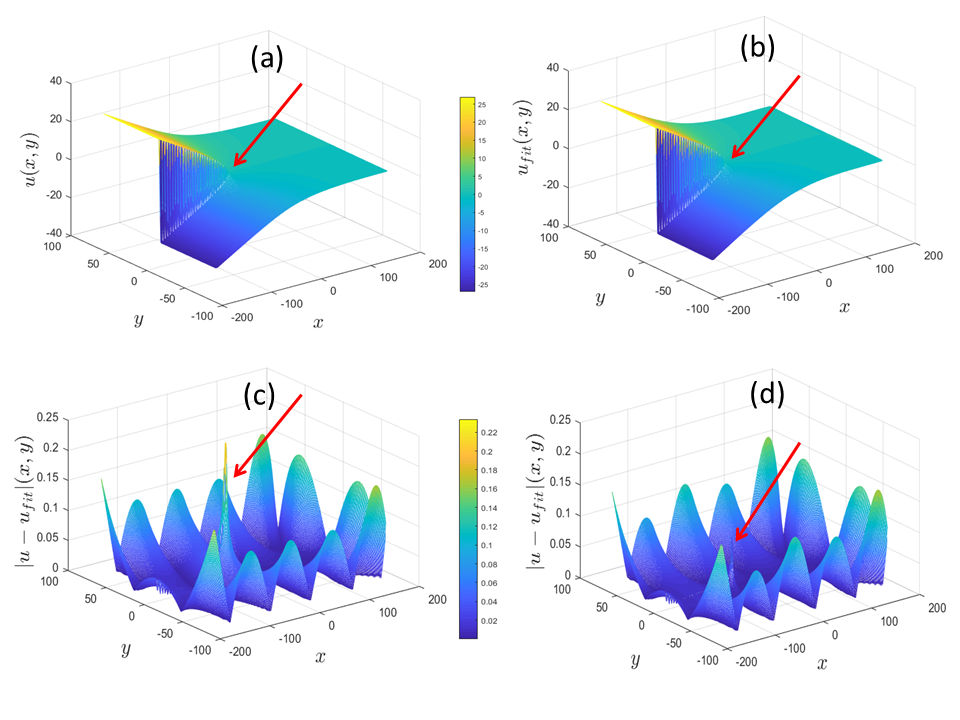}
\caption{Illustration of the procedure to position the continuum-level crack tip into discrete lattice: (a) voltage field calculated by simulation, $u(x,y)$, in a $400 \times 200$ honeycomb lattice under CT loading. {\bf b} fitted voltage field, $u_{fit} (x,y)$ using Eqs. \ref{eqWilliams} and \ref{eqWilliamsScal} (truncated to and with $n \leq 9$) with a tip origin (crack tip position) located in the middle $\OO$ of the next bond to break. {\bf c} absolute difference $|u_{fit}-u|(x,y)$ between the simulated and fitted voltage field shown in panels  {\bf a} and {\bf b} respectively. The fit is very good everywhere, except in the very vicinity of the next bond about to break (red arrow). {\bf d} absolute difference, $|u_{fit}-u|(x,y)$ after
having shifted the crack tip to the new position $\CC$ using the iterative procedure depicted in the text. The fit is now very good everywhere, even in the very vicinity of the next bond to break.  All quantities in the panels are expressed in reduced units: $\ell$ unit for positions $x$ and $y$,  $\sqrt{\gamma/g}$ unit for voltages $u(x, y)$ and $u_{fit} (x,y)$} 
\label{fig:2}
\end{figure*}

Figure \ref{fig:2}a presents the spatial distribution of the voltage field in a representative simulation. This field is the analog of the displacement field in a cracked slab under antiplanar loading conditions. As such, this field can be analyzed within the LEFM framework. 

\subsection{Williams expansion}
 
\cite{Williams52_jam} showed that, in a 2D elastic plate embedding a straight crack, the displacement field can be expressed as a series of elementary solutions: 

\begin{equation}
u=\sum_{n\geq 0} ^\infty a_n \Phi_n(r,\theta),
\label{eqWilliams}
\end{equation}

\noindent where $(r,\theta)$ are the polar coordinates of the considered point in a frame located at the crack tip. The elementary functions $\phi_n(r,\theta)$ depend on the fracture mode. The antiplanar elasticity examined here maps to a mode III fracture problem and $\phi^{III}(r,\theta)$ writes:

\begin{align}
\begin{split}
\Phi^{III} _n(r,\theta) & = r^{n/2}\sin{\frac{n\theta}{2}} \quad \mathrm{for} \; \mathrm{odd} \; n\\
                                      & = 0 \quad\quad\quad\quad  \mathrm{otherwise}
\label{eqWilliamsScal}
\end{split}
\end{align}

\noindent Note that LEFM discards super-singular terms ( $n<0$ terms in Eq. \ref{eqWilliams}), so as to ensure the finiteness of strain energy in the near-tip region. The $n = 1$ term coincides with the standard square-root singular term. The prefactor $a_1$ relates to stress intensity factor $K$ via: 

\begin{equation}
K = a_1 \frac{E\sqrt{2\pi}}{4}, \\
\label{eqK}
\end{equation}

\subsection{Tip positioning, fracture toughness, and non-singular terms determination}
\label{sec_FittingVotageField}

\begin{figure*}[htp]
\includegraphics[width =0.9\textwidth]{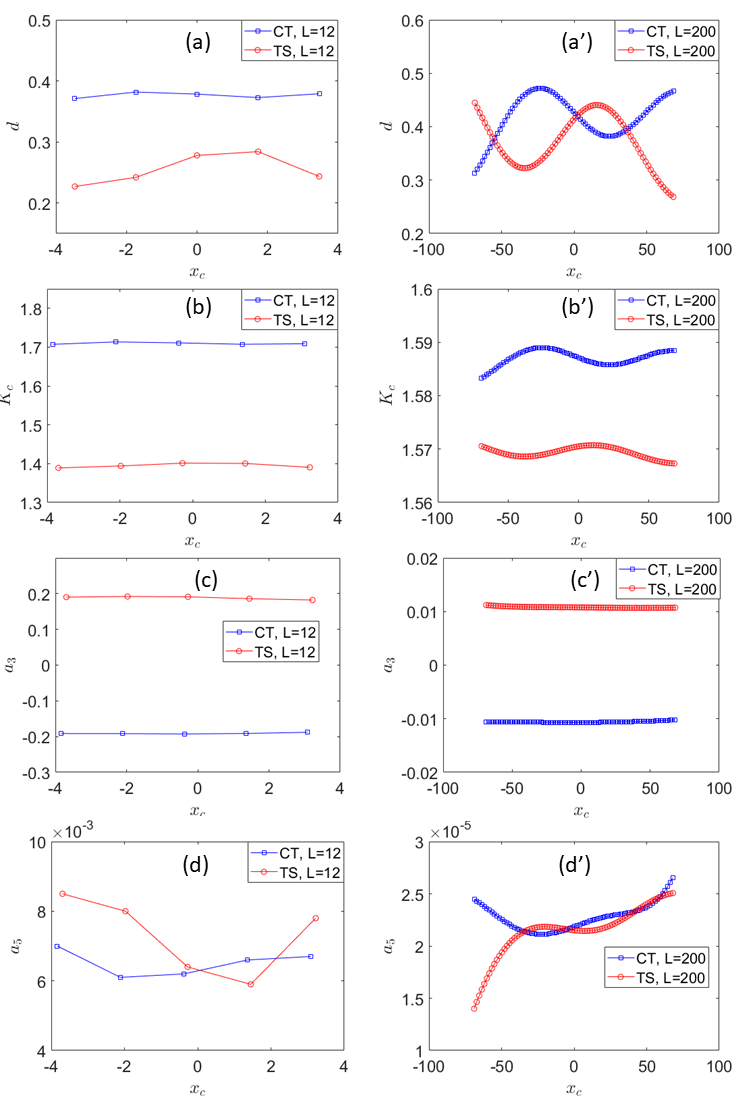}
\caption{ 
Variation of tip mispositioning $d$ (panels {\bf a} $\&$ {\bf a'}), fracture toughness $K_c$ (panels {\bf b} \& {\bf b'}), prefactor $a_3$ of first super-singular ($n=3$) term in Williams series (Eqs \ref{eqWilliams} and \ref{eqWilliamsScal}) (panels {\bf c} \& {\bf c'}) and prefactor $a_5$ of second super-singular ($n=5$) term in Williams series  (panels {\bf d} \& {\bf d'}). 
Measurements were carried out in a honeycomb lattice of sizes $L=12$ (smallest studied, left column,  {\bf a}$\rightarrow$  {\bf d}) and $L=200$ (largest studied, right column,  {\bf a'}$\rightarrow$  {\bf d'}). All quantities are plotted as a function of the position of the next fuse to break, $x_c$. All quantities are expressed in reduced units: $\ell$ unit for $x_c$, $d$ and $L$,  $\sqrt{g\gamma/\ell}$ unit for $K_c$, $\sqrt{g\gamma}/\ell^{3/2}$ and $\sqrt{g\gamma}/\ell^{5/2}$ units for $a_3$ and $a_5$.
}
\label{fig:d_Kc_a3_a5}
\end{figure*}

{\em A priori}, fitting numerical displacement/voltage fields with Williams' series (Eqs. \ref{eqWilliams} and \ref{eqWilliamsScal}) at the onset of bond breaking allows the determination of $K_c$.  All the difficulty is to properly place the crack tip in a discrete lattice. As a first guess, we placed it at the center of the next bond to break. Then,  we used Eq. \ref{eqWilliams}, truncated to the first five terms ($n \leq 10$) to fit $u(x,y)$ obtained in the numerical simulations. Figure \ref{fig:2}(b) presents the fit, $u_{fit}(x,y)$, associated to the numerical field $u(x,y)$ showed in Fig. \ref{fig:2}(a). Note the apparent similarity between Figs. \ref{fig:2}(a) and \ref{fig:2}(b). To compare them more quantitatively, we show in Fig. \ref{fig:2}(c) the absolute difference $|u_{fit}-u|(x,y)$. The fit is very good everywhere, except in the very vicinity of the crack tip. Unfortunately, this near-tip zone is precisely the one setting whether or not the next bond breaks.

As shown by \cite{Nguyen19_prl}, this near-tip discrepancy results from a mispositioning of the crack tip. As first shown by \cite{Rethore13_jmps} for the analysis of cracked samples in experimental mechanics, a slight mispositioning of the crack tip leads to the appearance of an additional super-singular ($n=-1$) term in Eq. \ref{eqWilliams}, with a prefactor $a_{-1}=a_1 d/2$ where $d$ is the mispositioning along $x$. An iterative procedure was then proposed  to find the proper tip position in a discrete fuse lattice \citep{Nguyen19_prl};
\begin{itemize}
\item[(i)] Start with a crack origin $\OO$ located at the center of the next bond to break;
\item[(ii)] Express the displacement field observed on the considered simulation snapshot in cylindrical coordinates for a frame centered on $\OO$
\item[(iii)] Fit $u(r,\theta)$ with Eq. \ref{eqWilliams} to which the super-singular $n=-1$ term has been added;
\item[(iv)] Define the new crack tip position (now $\CC$) so that it is horizontally shifted with respect to $\OO$ from $d=2 a_{-1}/ a_1$; 
\item[(v)] Repeat step (ii) to (iv) till the horizontal shift becomes negligible (in practice $d \leq 0.01$) so as the crack tip position has converged.
\end{itemize}

\noindent In all our simulations, convergence was achieved in less than three iterations, regardless of lattice geometry, lattice size, or loading scheme. The obtained fitted voltage field is now very good everywhere, even at the edge of the next bond to break (Fig. \ref{fig:2}(d)).

Note that Williams' expression (Eqs. \ref{eqWilliams} and \ref{eqWilliamsScal}) can be extended to address the stress field, and that it is then possible to derive an equivalent procedure to fit the local (virial) stress field obtained numerically; the procedure is detailed in Appendix 1. However, this is less convenient for two reasons: Firstly, the stress field has two non-zero components in antiplanar elasticity, whereas the displacement field is fully given by the (scalar) voltage field $u(x, y)$; secondly, the stress field diverges at the crack tip, which makes the procedure less sensitive for accurate detection of the tip position, and hence for determining the coefficients $\{a_n\}$. Hence, in the following, therefore, the analysis will be restricted to $u(x,y)$ only. 

Figures \ref{fig:d_Kc_a3_a5}(a) and \ref{fig:d_Kc_a3_a5}(a') show the profiles $d(x_c)$ of  mispositioning observed in honeycomb lattices for two different sizes: $L=12$ (Fig. \ref{fig:d_Kc_a3_a5}(a)) and $L=200$ (Fig. \ref{fig:d_Kc_a3_a5}(a')). Here $x_c$ refers to the position of the next bond to break. In both cases, $d$ depends very little on the position of the crack tip; this has been observed in all our simulations regardless of the lattice geometry and size. On the other hand, $d$ depends on the applied loading for small system sizes (around $50\%$ higher for CT loading than for TS loading), and becomes load-independent when $L$ is large. This behavior, too, was observed irrespective of lattice geometry and will be characterized in more detail in section 5.

Once the mispositioning is corrected and the crack tip is placed at its proper location $\CC$, the fracture toughness $K_c$ is determined using Eq. \ref{eqK} from the prefactor $a_1$ of the $n=1$ term of the fitted Williams series. Figures \ref{fig:d_Kc_a3_a5}(b) and \ref{fig:d_Kc_a3_a5}(b') show its evolution as a function of crack length in honeycomb lattices with  $L=12$ and $L=200$. As $d$, $K_c$ is almost independent of  $x_c$, and becomes independent of loading conditions at large $L$: Note the very small range over which $K_c$ varies for $L=200$, from $1.57$ to $1.59$: $K_c$ is also independent of loading conditions. The same has been observed for triangular and square lattice geometry.

Finally, to complete the characterization of the displacement field, we examine the prefactors $a_3$ and $a_5$ associated with the first two super-singular terms in Williams' expansion (Eq. \ref{eqWilliams}). The profiles observed in the honeycomb geometry are presented in Figs. \ref{fig:d_Kc_a3_a5}(c) and \ref{fig:d_Kc_a3_a5}(d) for $L=12$, and in Figs. \ref{fig:d_Kc_a3_a5}(c') and \ref{fig:d_Kc_a3_a5}(d') for $L=200$. Like $d$ and $K_c$, $a_3$ is independent of $x_c$, but unlike them, $a_3$ is highly dependent on CT or TS loading conditions. $a_5$ depends little on loading and $x_c$ and its average value is highly dependent on the system size $L$.

In any case, for a given geometry and loading, these parameters, $d$, $K_c$, $a_3$, and $a_5$, are fairly constant once the crack has sufficiently propagated. From now, these parameters are considered to be independent of the crack length. We will return in Sec. 5 to the analysis of their dependency on system size.  

\section{Energy balance at global scale: Fracture energy}

At present, while efforts to determine \textit{ab-initio} the fracture energy from atomistic parameters have been only partially successful, standardized tests are widely used in experimental mechanics to measure this quantity on various materials. The first method, referred to as the compliance method \citep{Gordon78}, consists in measuring the fracture energy from the evolution of the overall lattice conductance with the position of the crack tip $x_c$. A second procedure, referred to as the virtual work method, consists in identifying the fracture energy with the loss of energy stored in the lattice just as the burnt fuse is withdrawn, keeping $U_{ext}$ constant. As the fracture energy is independent of the measurement method \citep{Nguyen19_prl}, only the virtual work method was applied in this section. 

\subsection{Fracture energy measurement}
\label{sec:3}

\begin{figure*}[htp]
\includegraphics[width=\textwidth]{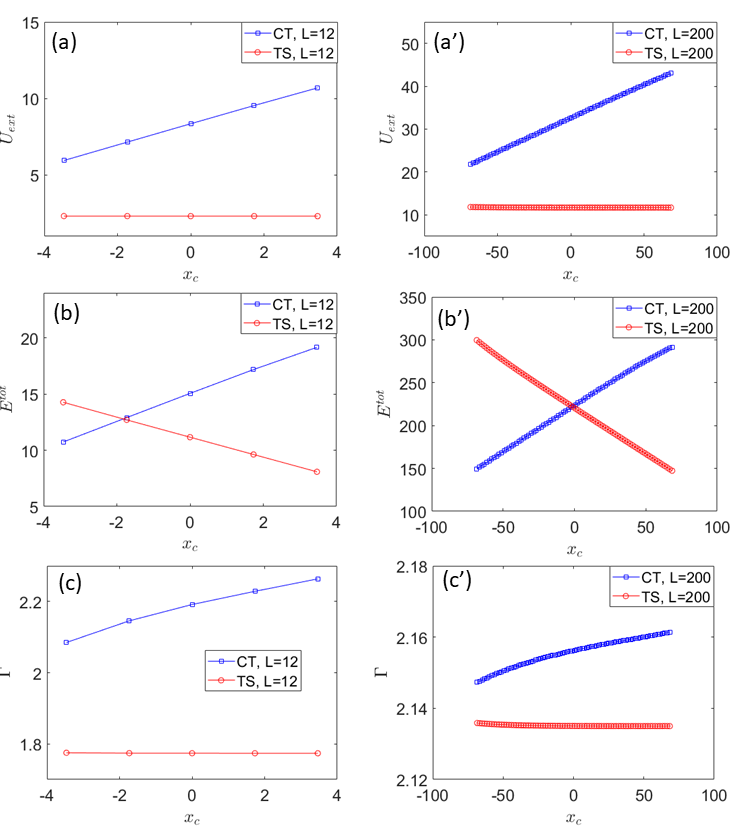}
\caption{
Illustration of the procedure used to determine fracture energy. Panels {\bf a} and {\bf a'}: Variation of the loading, $U_{ext}$, applied at the breakdown onset as a function of the position $x_c$ of the next bond to break. Under CT loading, $U_{ext}$ increases with $x_c$. Conversely, it is nearly independent when TS loading is applied. Panels {\bf b} and {\bf b'}: Profile of total energy $E^{tot}(x_c)$ stored in the lattice just before the breakdown event. Panels {\bf c} and {\bf c'}: Profile of fracture energy, defined as the decrease of total energy induced by the fuse breakdown divided by the crack length increment induced by this breakdown event (Eq. \ref{eq_Gamma_virtual_work}).Measurements were carried out in a honeycomb $2 L \times L$ lattice of sizes $L=12$ (smallest studied, left column,  {\bf a}$\rightarrow$  {\bf c}) and $L=200$ (largest studied, right column,  {\bf a'}$\rightarrow$  {\bf c'}). In all panels, blue squares correspond to CT loading while red circles correspond to TS loading. All quantities are expressed in reduced units: $\ell$ unit for $x_c$ and $L$,  $\sqrt{\gamma/g}$ unit for $U_{ext}$, $\gamma$ for $E^{tot}$, and $\gamma/\ell$ unit for $\Gamma$.
}
\label{fig:Uext_Etot_Gamma_Vs_xc}
\end{figure*}

 Using the virtual work method, we examine the loss of total energy stored in the lattice, $E^{tot}$, when the fuses are burnt and the crack propagates. $E^{tot}$ is defined as $E^{tot} = \sum_p i_p ^2/2g$ where $i_p$ is the current flowing through the fuse $p$, at the time considered. Fracture energy, $\Gamma$, is then defined as the variation of $E^{tot}$ between before and after the fuse burns (keeping  the applied loading $U_{ext}$ constant and equal to the value at the burning onset), divided by the crack length increment $\delta x_c$ caused by the fuse breakdown:  
 
 \begin{equation}
     \Gamma (x_c) = \frac{E^{tot}(x_c^-)-E^{tot}(x_c^+)}{\delta x_c}
      \label{eq_Gamma_virtual_work}
 \end{equation}

 Figure \ref{fig:Uext_Etot_Gamma_Vs_xc} illustrates the procedure. Figures \ref{fig:Uext_Etot_Gamma_Vs_xc}(a), \ref{fig:Uext_Etot_Gamma_Vs_xc}(a'), \ref{fig:Uext_Etot_Gamma_Vs_xc}(b) and \ref{fig:Uext_Etot_Gamma_Vs_xc}(b') present the variations of $U_{ext}$ and $E^{tot}$ at breaking onset with $x_c$, respectively, in  honeycomb lattices of small and large size, with either CT or TS loading applied. Qualitatively similar variations are observed in all the lattices, regardless of their size and geometry. Note that the profiles are highly dependent on the prescribed loading configuration: Both $U_{ext}$ and $E^{tot}$ increase with crack length when CT loading is prescribed, while $U_{ext}$ remains fairly constant and $E^{tot}$ decreases with $x_c$ for TS loading. 
 
Figures \ref{fig:Uext_Etot_Gamma_Vs_xc}(c) and \ref{fig:Uext_Etot_Gamma_Vs_xc}(c') show the variations of $\Gamma$ as determined via Eq. \ref{eq_Gamma_virtual_work} in the same honeycomb lattices as that associated with Figs. \ref{fig:Uext_Etot_Gamma_Vs_xc}(a) and \ref{fig:Uext_Etot_Gamma_Vs_xc}(b) (smallest size, $L=12$), and Figs. \ref{fig:Uext_Etot_Gamma_Vs_xc}(a') and \ref{fig:Uext_Etot_Gamma_Vs_xc}(b') (largest size, $L=200$). In both cases, $\Gamma$ is nearly independent of crack length and weakly dependent on the loading conditions. The difference measured on $\Gamma$ between TS and CT loading is significant at the smallest size ($\sim 20\%$ for $L=12$), and vanishes as system size becomes very large ($\sim 1\%$ for $L=200$). 

\subsection{Verification of Irwin's relation}
\label{sec:Irwin_verification}

\begin{figure}[htp]
\centering
\includegraphics[width=0.5\textwidth]{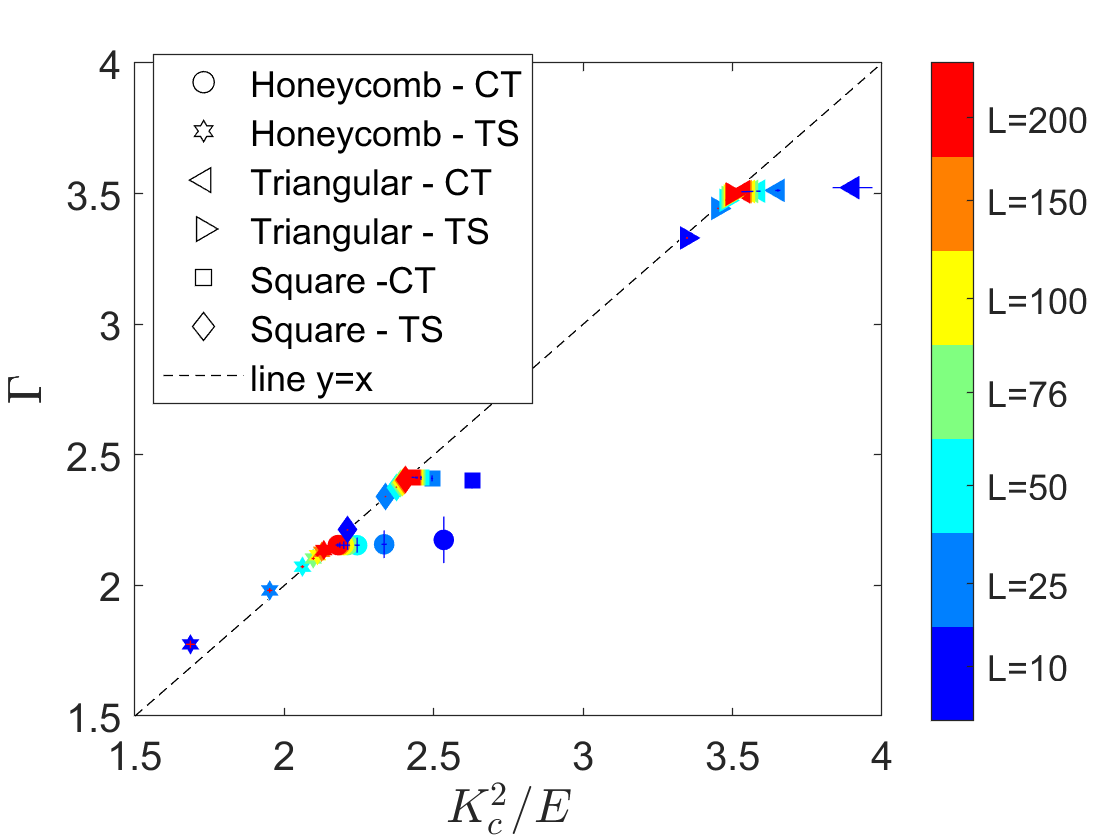}
\caption{Fracture energy, $\Gamma$, as a function of $K_c^2/E$ in honeycomb, triangular, and square lattices of different sizes, under either CT or TS loading. Different symbols are used as markers to indicate lattice geometry and imposed loading. Their color indicates the lattice size according to the colorbar on the right (dark blue for the smallest lattice, $L=10$ to red for the largest one, $L=200$. In all simulations, $\Gamma$ and $K_c$ are averaged over the crack tip position. A straight dotted line shows $y=x$ as expected from Irwin's relation (Eq. \ref{eqIrwin}).$L$ is expressed in $\ell$ unit,  and $\Gamma$ and  $K_c^2/E$ are expressed in  $\gamma/\ell$ unit.}
\label{fig_GammaSim_Vs_GammaIrwin}
\end{figure}

We now compare, in our simulation, the resistance-to-failure determined from the displacement field, $K_c$, and that determined at the global scale from Griffith's energy balance, $\Gamma$. Within the LEFM framework, the two are related by \citep{Irwin57_jam}:

\begin{equation}
\Gamma = \frac{K^2_c}{E},
\label{eqIrwin}
\end{equation}

As illustrated in Figs. \ref{fig:d_Kc_a3_a5}(b) and \ref{fig:Uext_Etot_Gamma_Vs_xc}(c), both $K_c$ and $\Gamma$ are independent of the crack tip position $x_c$; hence, here and in the following, we define $\overline{\Gamma}=\langle \Gamma(x_c) \rangle_{x_c}$ and $\overline{K}_c=\langle K_c(x_c) \rangle_{x_c}$ as the values averaged over $x_c$. Moreover, we drop the bar over the averaged quantities for the sake of simplicity. 

Figure \ref{fig_GammaSim_Vs_GammaIrwin} tests Irwin's relation and shows $\Gamma$ as a function of $K^2_c/E$. Each point corresponds to a given simulation with a lattice size ranging from $L=10$ to $L=200$, honeycomb, triangular or square geometry, and either CT or TS loading configuration. Regardless of the lattice geometry and loading condition,  Eq. \ref{eqIrwin} is perfectly satisfied as soon as the lattice size is large enough, i.e. $L \geq 50$. A discrepancy is observed at smaller system sizes. They are characterized in the next section. 

\section{On the system-size dependency}

In this section, we investigate the effect of specimen size on the different parameters calculated previously.

\subsection{Mis-positioning}
\label{sec:7}

\begin{figure}[htp]
\includegraphics[width=0.5\textwidth]{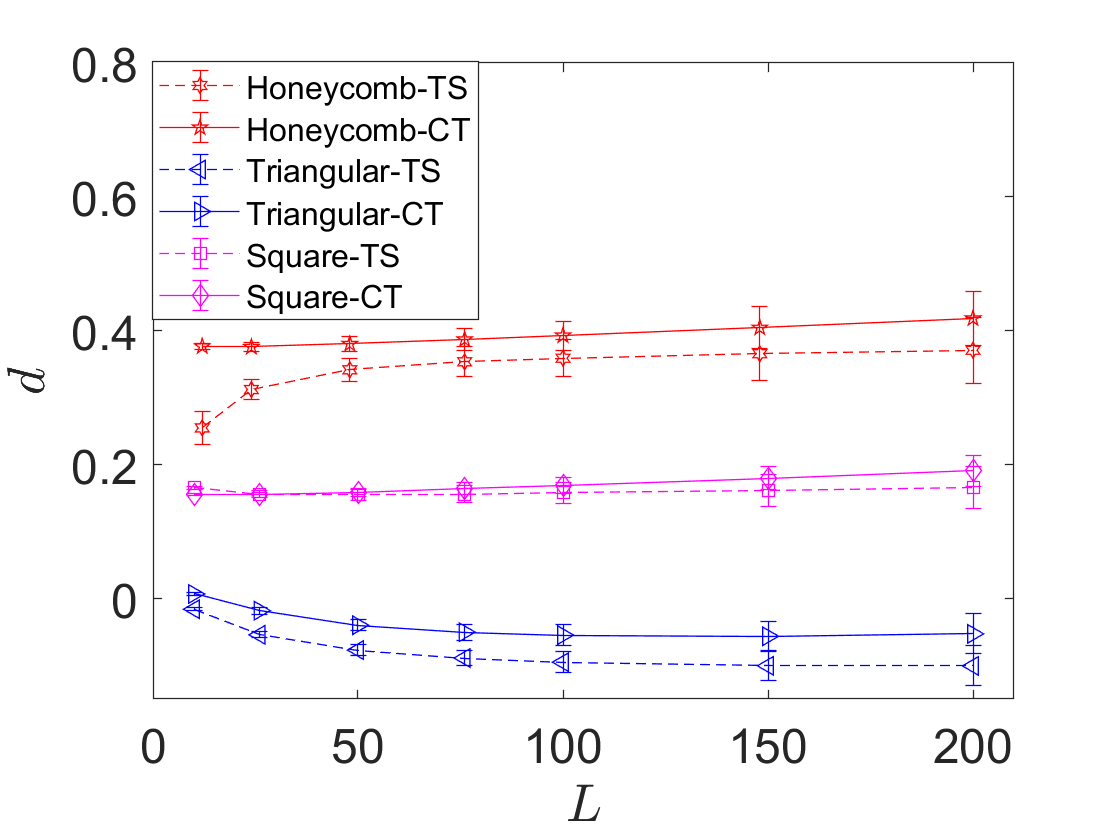}
\caption{Mis-positionning $d$  as a function of system size $L$ on the three lattice geometries under either CT or TS loading configuration. Each point corresponds to a given simulation and $d(x_c)$ was averaged over crack tip position $x_c$. The different symbols and their significance are provided in the legend on the left. $L$ and $d$ are expressed in $\ell$ unit.} 
\label{fig:d_Vs_L}
\end{figure}

As for $K_c$ and $\Gamma$, we now consider the $x_c$-averaged value $\overline{d}=\langle d(x_c) \rangle_{x_c}$ and, in the following, drop the bar in the notation for the sake of simplicity. The dependence of $d$ on the system size $L$ is now represented in Fig. \ref{fig:d_Vs_L}. $d$ can be quite large, up to $\sim 40\%$ of the bond length in honeycomb lattices. It is positive in honeycomb and square lattices, which means that the true tip position is before the center of mass of the next bond to break. Conversely, $d$ is negative in triangular lattices, which means that the true continuum-level scale tip position lies in the part of the lattice yet to be broken, above the next bond to break. In addition, $d$ weakly depends on the loading configuration value; it is always smaller in the TS loading configuration than in the CT one. Finally, $d$ tends toward a value almost independent of $L$ when $L$ is larger than 100. In the following, we define the mispositioning in the continuum limit, $d^{\infty}$, as the average of $d(L)$ over $L \geq 100$ for both CT and TS loading. The obtained values are reported in the first line of Tab. \ref{tab_d}.

\subsection{Fracture toughness and sub-singular terms}
\label{sec:frac_t} 

\begin{figure}[htp]
\centering
\includegraphics[width=0.5\textwidth]{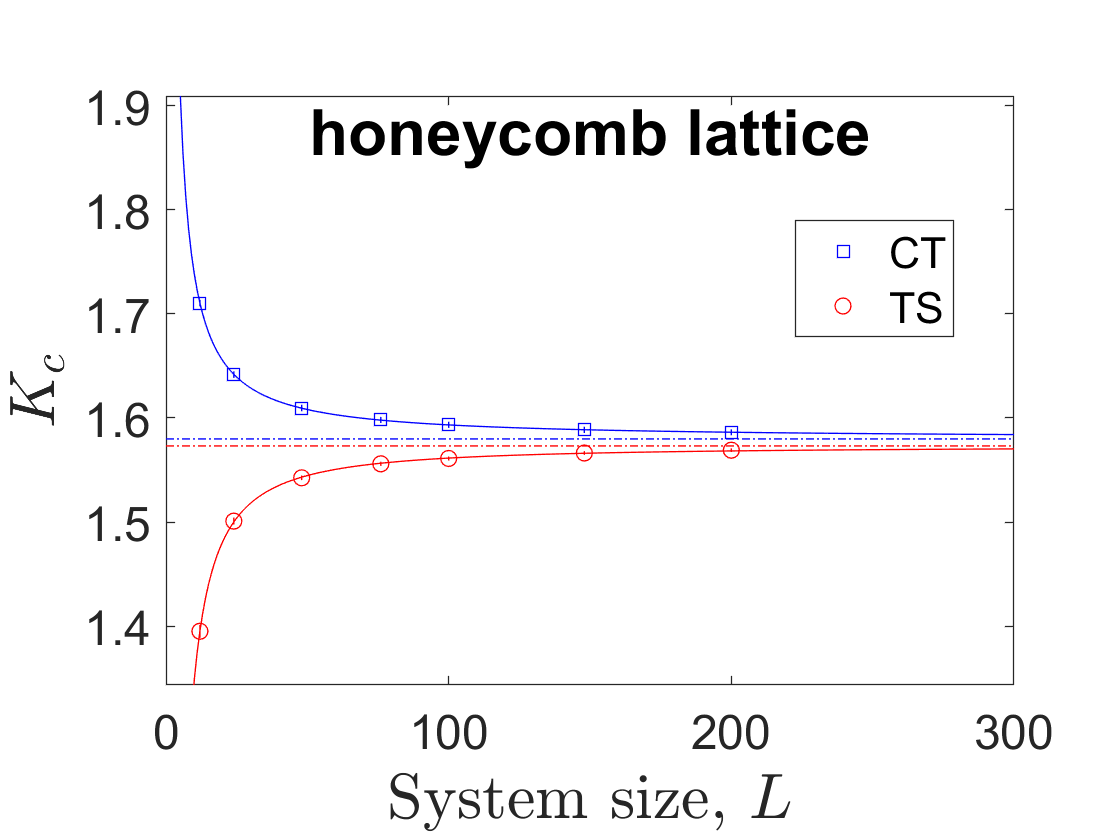}
\includegraphics[width=0.5\textwidth]{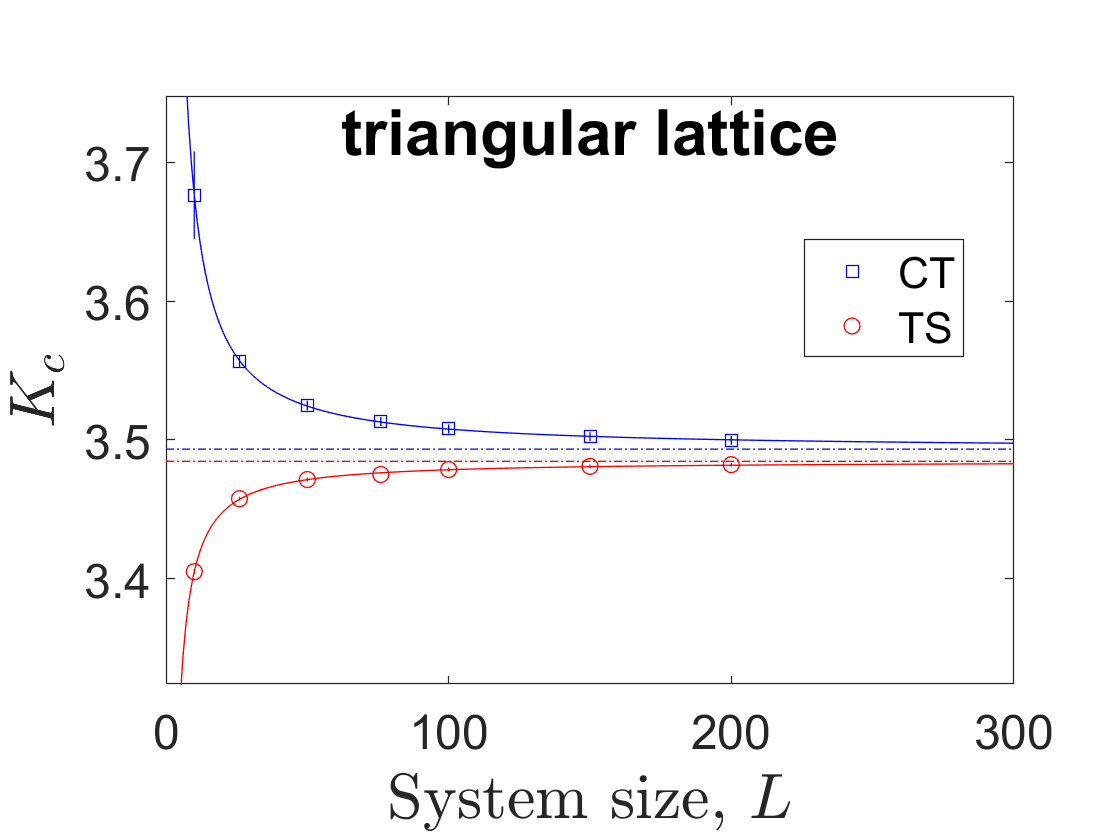}
\includegraphics[width=0.5\textwidth]{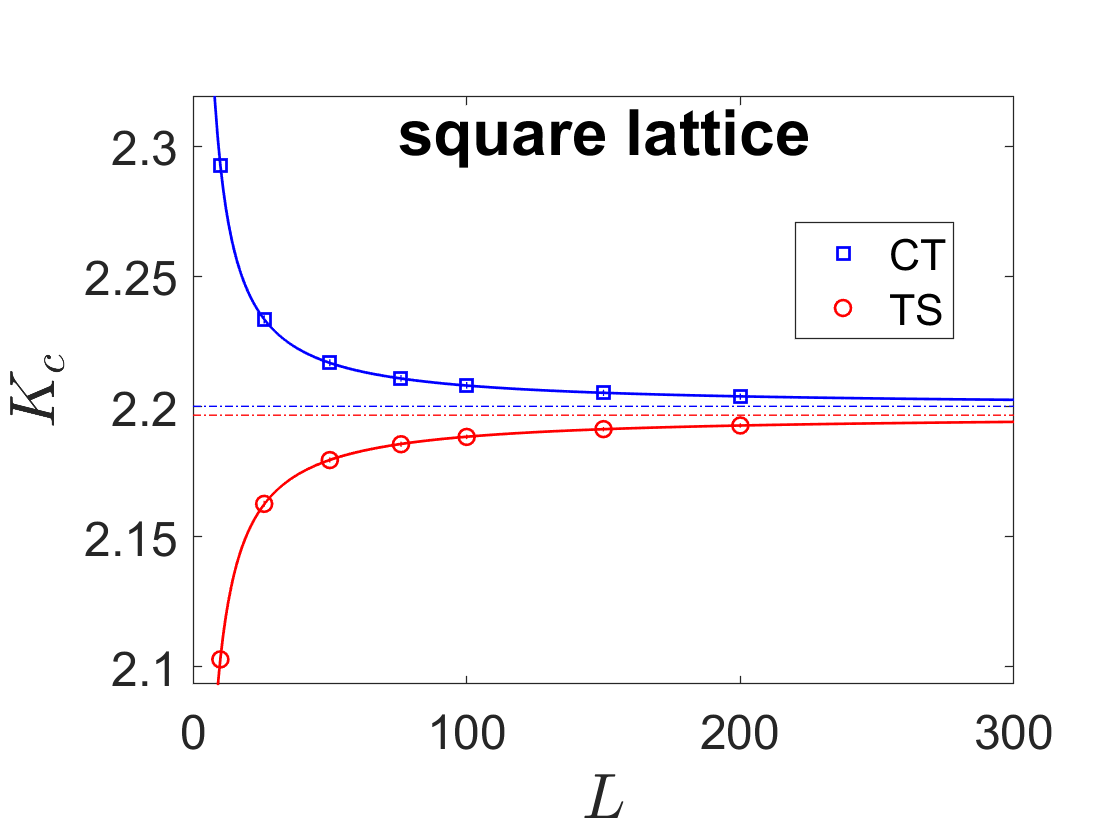}
\caption{Fracture toughness, $K_c$ as a function of system size $L$ for the different lattice geometries: Honeycomb on the top panel, triangular on the middle panel, and square on the bottom one. Blue circle and red square symbols on each panel correspond to CT and TS loading, respectively. Each point corresponds to a given simulation and $K(x_c)$ was averaged over crack tip position $x_c$. Solid blue and red lines are fits via $K_c(L)=a / L^p + K^{\infty}_c$. Blue and red horizontal dotted lines indicate $K^{\infty}_c$ for CT and TS loading, respectively. $L$ is expressed in $\ell$ unit,  and $K_c$ is expressed in  $\sqrt{g\gamma/\ell}$ unit. Adapted from \cite{Nguyen19_prl}.}
\label{fig:Kc}
\end{figure}

 \begin{figure}[htp]
 \includegraphics[width=0.5\textwidth]{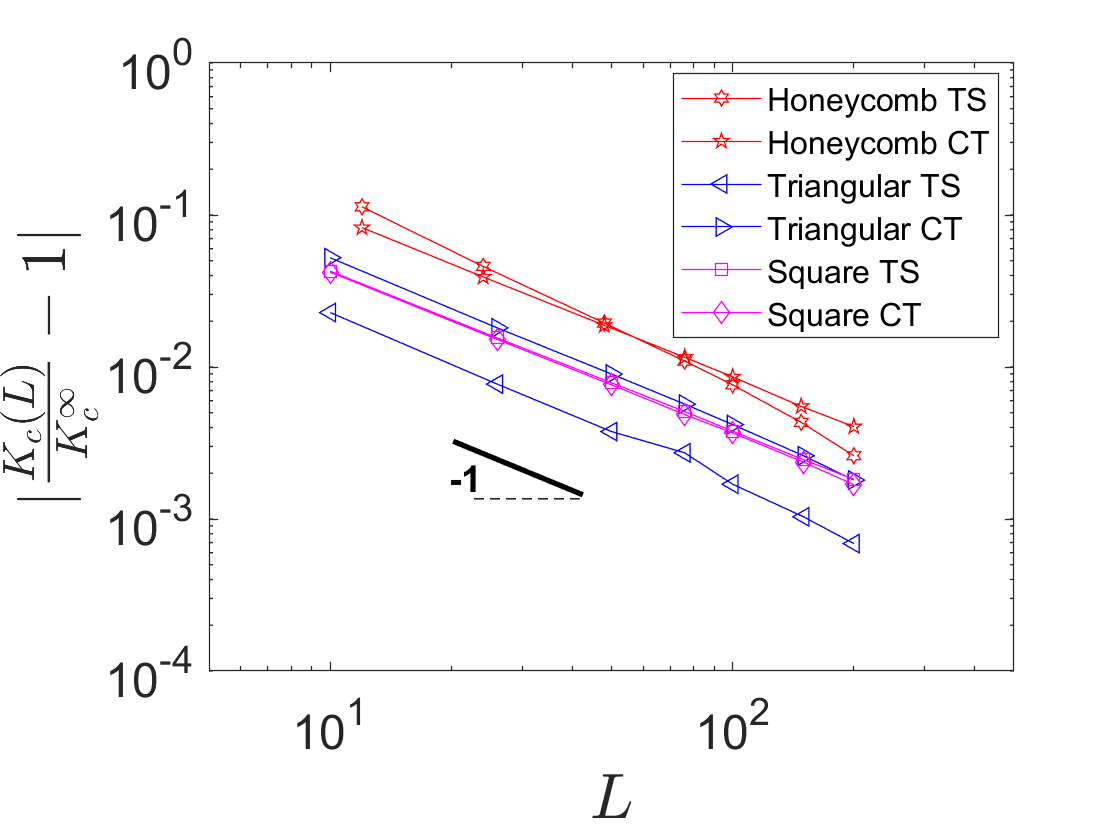}\hfill
\includegraphics[width=0.5\textwidth]{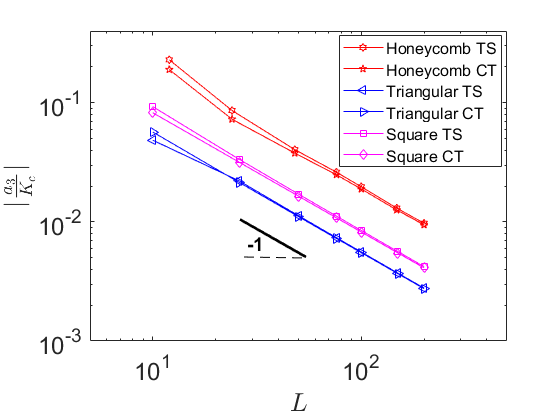}
\includegraphics[width=0.5\textwidth]{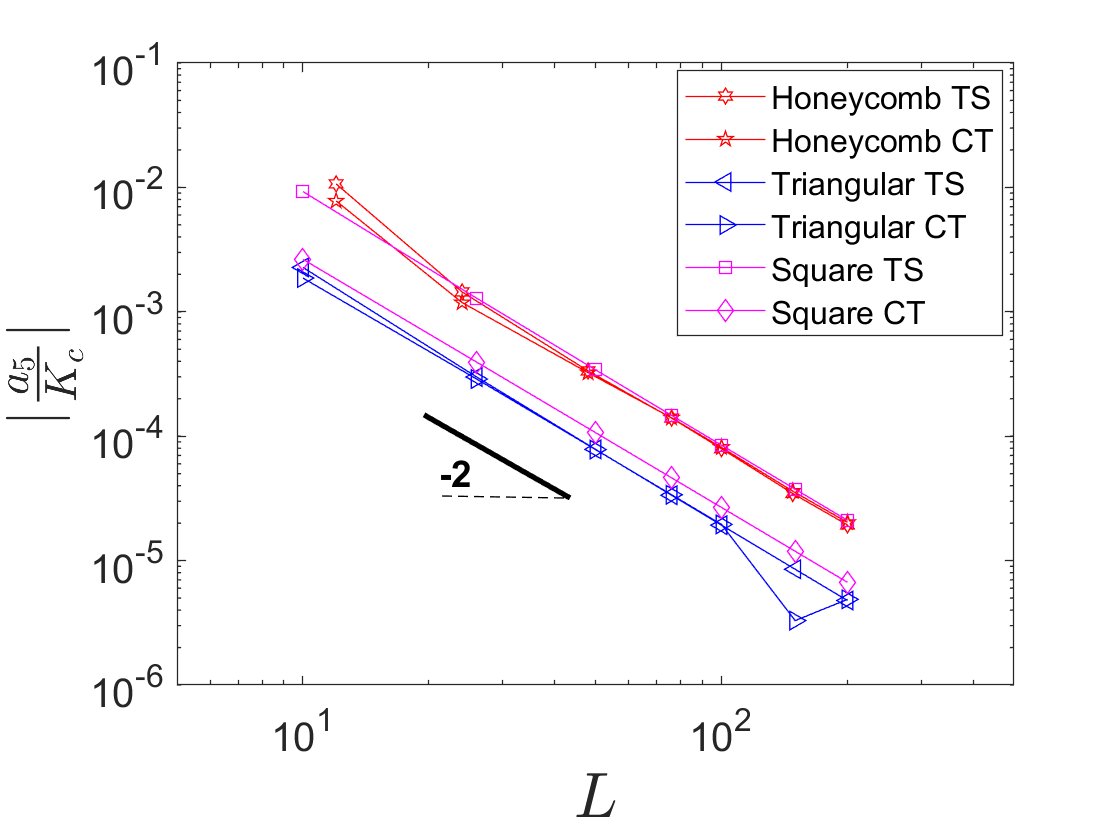}
\caption{First panel: the convergence of $K_c$ toward the continuum limit, $K_c^\infty$, as a function of system size $L$ in the honeycomb, triangular and square lattices under either CT or TS loading (different symbols). In all cases, the associated exponent close to $-1$ is shown by the black solid line. Second and third panels: $a_3/K_c$ and $a_5/K_c$ as a function of $L$. In these log-log scales, the black solid line shows the slope of $-1$ in the second panel and the slope of $-2$ in the third, as expected from Eq. \ref{eqWilliams}. Each point correspond to a given simulation by averaging $K_c(x_c)$, $a_3(x_c)$ and $a_5(x_c)$  over crack tip position. $L$ is expressed in $\ell$ unit.}
\label{fig:8}
\end{figure} 

Figure \ref{fig:Kc} shows the variations of fracture toughness with system size in honeycomb, square, and triangular lattices submitted to either CT or TS loading. In all cases, the $K_c\, vs. \, L$ curves are well fitted by $K_c = a/L^p+K_c^{\infty}$ (fitted parameters reported in Tab. \ref{tabKc}). Provided that the lattice geometry is prescribed, the fitted values $K_c^\infty$ determined for CT and TS loading are very close. The average between the two is reported in the last column of Tab. \ref{tabKc}, and defines the continuum-level scale fracture toughness, which, due to its independence with respect to loading conditions, is a material constant.

The first panel of Fig. \ref{fig:8} shows how $K_c$ converges toward $K_c^{\infty}$ for the different geometries and loading conditions. In all cases, the convergence is algebraic, with an exponent $p$ close to $1$ (Tab. \ref{tabKc}, fourth column). This power-law behavior implies the absence of characteristic length-scale in the specimen-size dependency. On average, the prefactor of the power-law scaling is the largest in the honeycomb lattice and the smallest in the triangular one, suggesting that the higher the lattice symmetry, the faster the convergence toward the continuum-level scale limit.

We now examine the contribution of the first two sub-singular terms of Williams expansion; these are given by the two prefactors $a_3$ and $a_5$ in Eq. \ref{eqWilliamsScal}. As for $K_c$, $a_3(x_c)$ and $a_5(x_c)$ are first averaged over crack tip position $x_c$ for each simulation. Figure \ref{fig:8} presents the ratios ${a_3}/{K_c}$ and ${a_5}/{K_c}$ as a function of system size $L$. The second ratio is about two orders of magnitude smaller than the first one. They both decrease as a power law with $L$, with an exponent $-2$ for $a_5/K_c$ and $-1$ for $a_3/K_c$, regardless of the lattice geometry and loading configuration. These exponents can be understood from dimensional analysis. The prefactors $a_n$ (n=1,3,5,...) in Eq. \ref{eqWilliams} are expressed in terms of $[\mathrm{length}]^{1-n/2}$. Since all dimensions are set by $L$, we expect $a_n \sim L^{1-n/2}$, hence ratios $a_n/K_c \sim 1/L^{(n-1)/2}$. When $L \rightarrow \infty$, both $a_3$ and $a_5$ (and more generally all $a_n$ with $n > 1$) are negligible with respect to $K_c$ and can be ignored.

Note that if the scaling exponents are roughly independent of lattice geometry, the convergence rate significantly depends on it. For both $a_3$ and $a_5$ and as for $K_c$, it decreases with the lattice symmetry (highest prefactor in honeycomb, smallest in triangular lattice). Note that this convergence rate is almost independent of the loading conditions. 

\subsection{Fracture energy}
\label{sec:8}

\begin{figure}[htp]
\includegraphics[width=0.5\textwidth]{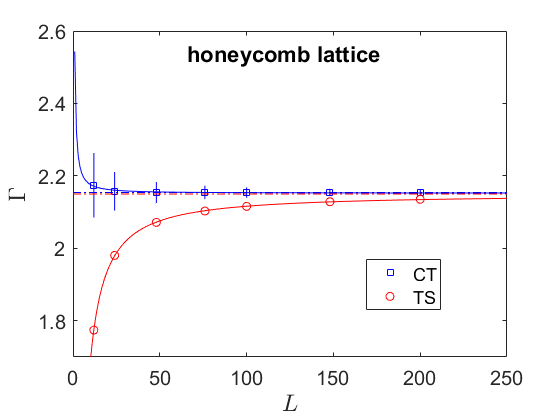}\hfill
\includegraphics[width=0.5\textwidth]{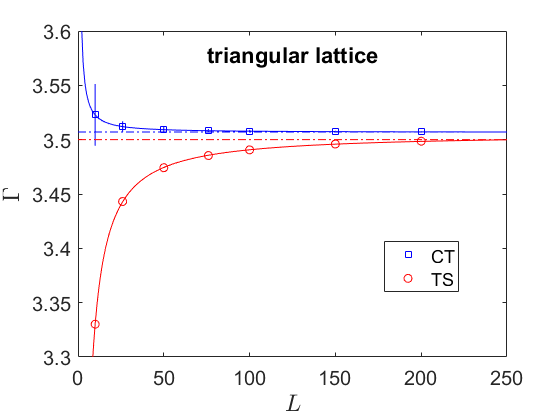}
\includegraphics[width=0.5\textwidth]{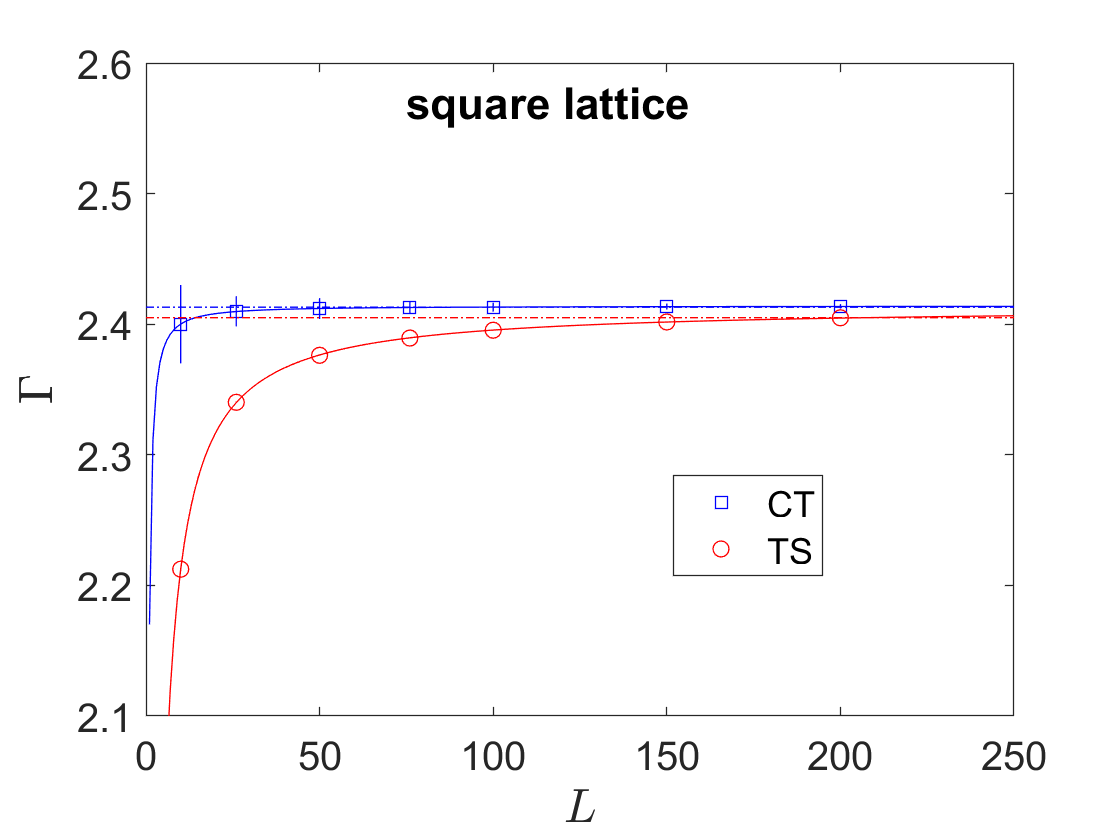}
%\end{center}
\caption{Fracture energy, $\Gamma$ as a function of system size $L$ for the different lattice geometry: Honeycomb on the top panel, triangular on the middle panel, and square on the bottom one. Each point corresponds to a given simulation and $\Gamma(x_c)$ was averaged over crack tip position $x_c$. Blue square and red circle symbols on each panel correspond to CT and TS loading, respectively. Solid blue and red lines are fits via the power law $\Gamma (L)= \alpha / L^q + \Gamma^{\infty}$. Blue and red horizontal dotted lines indicate $\Gamma^{\infty}$ for CT and TS loading, respectively. $L$ is expressed in $\ell$ unit,  and $\Gamma$ is expressed in  $\gamma/\ell$ unit. Panel on the bottom has been adapted from \cite{Nguyen19_prl}.} 
\label{fig:FractureEnergy_Vs_L}
\end{figure}

\begin{figure}[htp]
\includegraphics[width=0.5\textwidth]{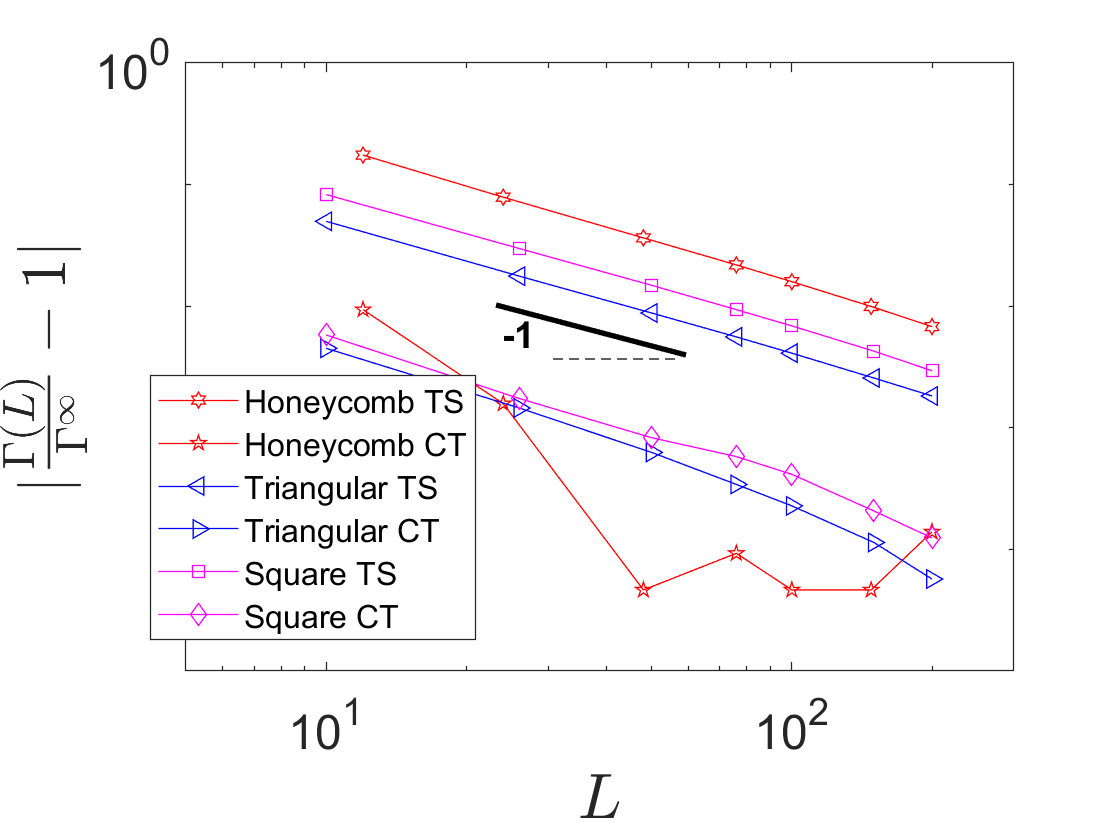}
\caption{$\frac{\Gamma(L)}{\Gamma^\infty} - 1 $, as a function of $L$ on the three lattice geometries under either CT or TS loading. $L$ is expressed in $\ell$ unit.}
%\label{fig:9}
\label{fig:GammaSim_GammaInf}
\end{figure}

We now study the dependence of fracture energy with system size $L$ (Fig. \ref{fig:FractureEnergy_Vs_L}). All curves $\Gamma \, vs. \, L$ are well fitted by $\Gamma (L) = \alpha/L^q+\Gamma^{\infty}$ where the fitted parameters $\alpha$, $q$ and $\Gamma^{\infty}$ are reported in Tab. \ref{tab :coeff_powerlaw_Gamma}. We can distinguish two regimes:

\begin{itemize}
    \item For large $L$, $\Gamma$ converges toward a constant value, $\Gamma^{\infty}$. As $K_c^\infty$, this value depends on the specific lattice geometry (triangular, honeycomb, square), but is roughly independent of loading configuration. This value hence defines the material-constant continuum-level scale fracture energy.    
     \item For small $L$ ($L < 100$), the curve $\Gamma(L)$ depends on the loading configuration. $\Gamma$ is always an increasing function of $L$ and converges to $\Gamma^{\infty}$ from below when TS loading is applied. Regarding CT loading, $\Gamma$ is also an increasing function of $L$ in square lattice, but a decreasing function of $L$ in honeycomb and triangular lattices.
\end{itemize}

Figure \ref{fig:GammaSim_GammaInf} analyzes how fast $\Gamma(L)$ converges toward the continuum limit $\Gamma^\infty$. In all cases, $\Gamma(L)$ converges algebraically toward $\Gamma^\infty$. In all cases except one of the honeycomb lattices under CT loading, the exponent $q$ associated with the algebraic convergence is relatively close to unity. The convergence is slower for TS loading, and faster when lattice symmetry increases.

\begin{table*}[htp]
    \centering
\begin{tabular}{|l|c|c|c|}
  \hline
   & Honeycomb & Triangular & Square  \\
  \hline
   $d^{\infty}$ & $0.39 \pm 0.02$  &  $-0.08 \pm 0.02$   & $0.17 \pm 0.01$\\
  \hline
  $d_{theo}(1^{st}order)$ & $0.6158$ & $0.1140$ & $0.2978$ \\
   \hline
   $\Gamma^{\infty}$ & $2.152 \pm 0.004$  &  $3.504 \pm 0.008$   & $2.409 \pm 0.009$\\
  \hline
   $\Gamma_{theo}(1^{st} order)$ & $2.5557$ & $3.6734$ & $2.7640$ \\
   \hline
\end{tabular}
  \caption{Mis-positioning determined from the simulations in the limit $L \rightarrow \infty$ (first line) and by the 1$^{st}$ order analytical method (second line) presented in Sec. 6.1. Fracture energy calculated from simulation (third line) and by the 1$^{st}$ order analytical method (fourth line) presented in Sec. 6.1.  $d_{theo}$ is expressed in $\ell$ unit, and $\Gamma_{theo}$ is expressed in $\gamma/\ell$ unit.}
  \label{tab_d}
\end{table*}

\begin{table*}[htp]
\centering
\begin{tabular}{|c|c|c|c|c|}
\hline
Geometry & Loading & a & $p$ & $K_{c}^{\infty}$ \\
\hline
\multirow{2}{*}{Honeycomb} &  TS & $-10.550 \pm 1.662$ & $1.285 \pm 0.051 $ & \\
\cline{2-4}
 & CT & $3.900 \pm 0.238$ & $1.069 \pm 0.205 $ &\multirow{1}{*}{$1.577\pm 0.005$}\\
\hline
\multirow{2}{*}{Square}& TS &$-2.205 \pm 0.094$ & $1.054 \pm 0.015$ & \\
\cline{2-4} & CT & $2.255 \pm 0.071$ & $1.066 \pm 0.011$ &  \multirow{1}{*}{$2.199 \pm 0.002$} \\
\hline
\multirow{2}{*}{Triangular} & TS & $-2.251 \pm 0.686$ & $1.116\pm 0.107 $ & \\
\cline{2-4}
& CT & $4.974 \pm 0.404$& $1.103 \pm 0.029$ & \multirow{1}{*}{$3.489\pm 0.007$} \\
\hline
\end{tabular}
\caption{Summary of the $p$ and $K_c ^{\infty}$ values obtained from the fit of fracture toughness curves in Fig. \ref{fig:Kc} by power law $a/L^p + K_c ^{\infty}$. $L$ is expressed in $\ell$ unit, and $K_c ^{\infty}$ is expressed in $\sqrt{g \gamma/\ell}$ unit.}
\label{tabKc}
\end{table*}

\begin{table*}[htp]
\centering
\begin{tabular}{|c|c|c|c|c|}
\hline
Geometry & Loading & $\alpha$ & $q$ & $\Gamma^\infty$\\
\hline
\multirow{2}{*}{Honeycomb} &  TS & $-6.367\pm 0.474$ & $1.140 \pm 0.032 $ & \\
\cline{2-4}
 & CT & $14.620\pm 18.175 $ & $2.650 \pm 0.503 $ &\multirow{1}{*}{$2.152\pm 0.004$}\\
\hline
\multirow{2}{*}{Square}& TS &$-7.675 \pm 5.026$ & $1.486 \pm 0.269$ & \\
\cline{2-4} & CT & $-1.081 \pm 2.104$ & $1.767 \pm 0.415$ &  \multirow{1}{*}{$2.409\pm 0.009$} \\
\hline
\multirow{2}{*}{Triangular} & TS & $-6.364 \pm 4.375$ & $1.460\pm 0.283 $ & \\
\cline{2-4}
& CT & $0.760 \pm 0.447$ & $1.569 \pm 0.241$ & \multirow{1}{*}{$3.504\pm 0.008$} \\
\hline
\end{tabular}
\caption{Summary of the $\alpha$, $q$ and $\Gamma^{\infty}$ values obtained from the fit of fracture energy curves in Fig. \ref{fig:FractureEnergy_Vs_L} by power law $\alpha/L^q + \Gamma ^{\infty}$. $L$ is expressed in $\ell$ unit, and $\Gamma^{\infty}$ is expressed in $\gamma/\ell$ unit.}
\label{tab :coeff_powerlaw_Gamma}
\end{table*}

\section{Confrontation to analytical predictions}

\subsection{1st order analytical theory}

\begin{figure*}[htp]
\includegraphics[width=0.99\textwidth]{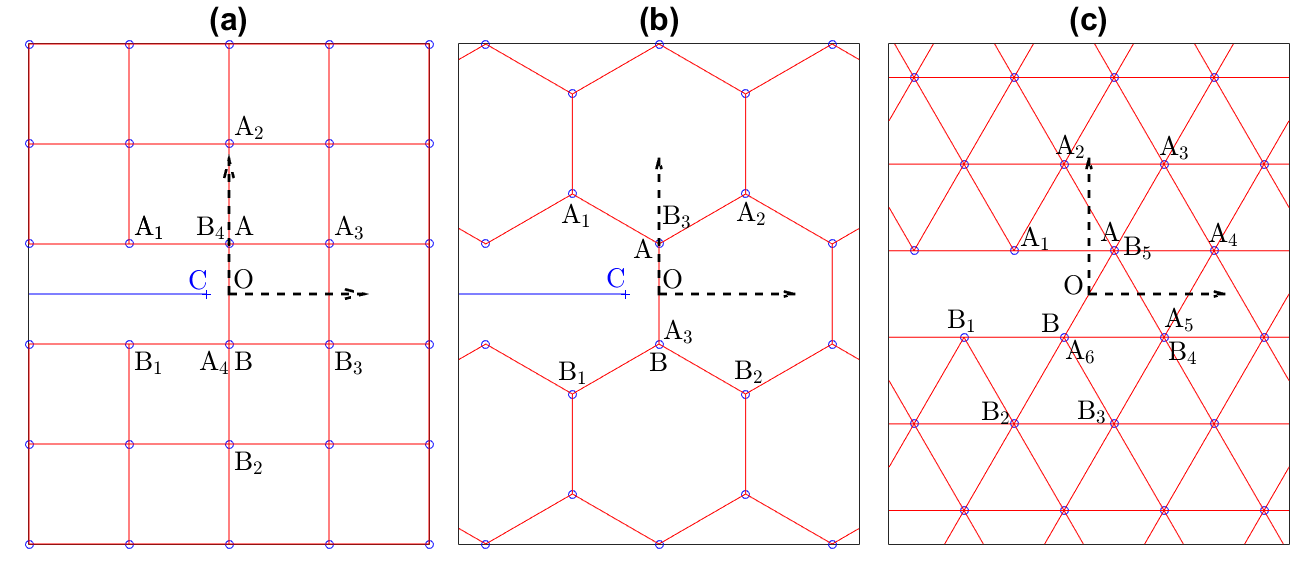}
\caption{Sketch and notation used in the analytical procedure to determine the continuum-level scale crack
tip and subsequently the fracture toughness as a function of
the lattice geometry, namely square (panel {\bf a}), honeycomb (panel {\bf b}) and triangular (panel {\bf c}).}
\label{fig:Sketch}
\end{figure*}

\begin{figure}[htp]
\includegraphics[width=0.5\textwidth]{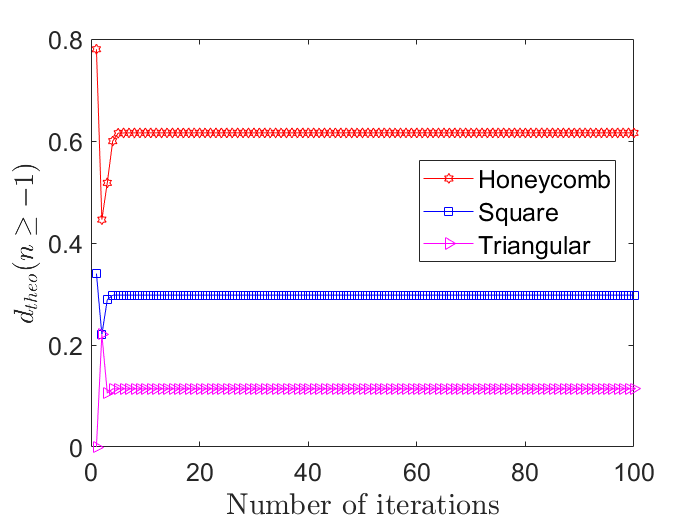}
\caption{Mis-positionning, $d_{theo}(n \geq -1)$ calculated via the $1^{st}$ order analytical method described in Sec. 6.1as a function of the number of iterations. The different symbols correspond to the different lattice geometries. In all cases, the lattice size is set to $L=200$.  $d_{theo}$ and $L$ are expressed in $\ell$ unit.}
\label{fig:d_Vs_Number_Itterration}
\end{figure}

\begin{figure}
\includegraphics[width=0.5\textwidth]{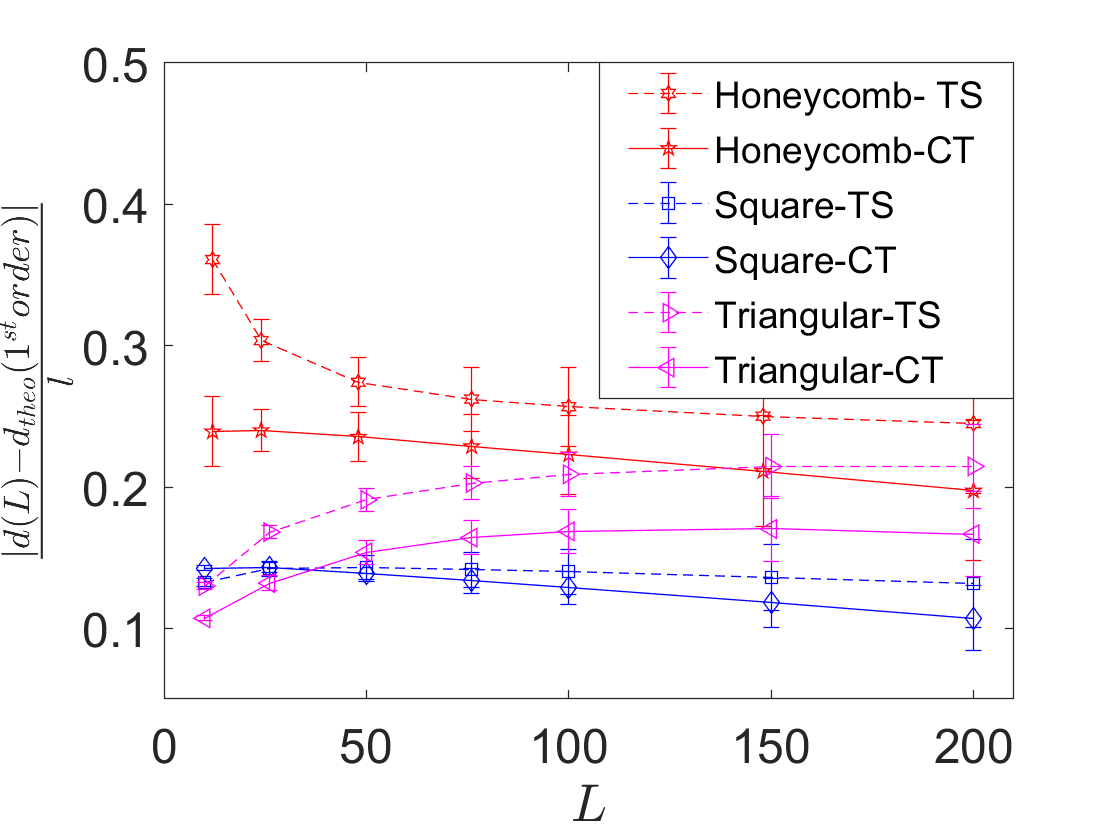}
\includegraphics[width=0.5\textwidth]{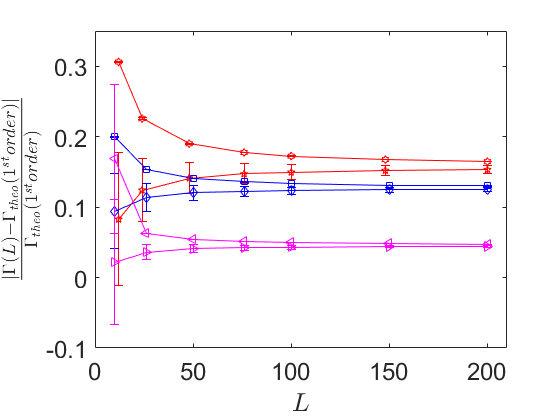}
\caption{Top: Difference between the mispositioning determined in the simulation, $d(L)$, and the one calculated by the 1$^{st}$ order analytical method, $d_{theo}(n \geq -1)$ (Tab. \ref{tab_d}, line 2) plotted as a function of the system size $L$. The difference is normalized by the bond length $l$. 
Bottom: Relative difference between the fracture energy determined in the simulation, $\Gamma(L)$, and the one calculated by 1$^{st}$ order analytical method, $\Gamma_{theo}(1^{st}order)$ (Tab. \ref{tab_d}, line 4) plotted as a function of the system size $L$. $L$ is expressed in $\ell$ unit.} 
\label{fig:13}
\end{figure}

\cite{Nguyen19_prl} proposed a simple analytical method to determine fracture energy from the atomistic parameters and the lattice geometry. To do so, they consider a lattice of infinite size and consider the displacement field at one of the two edges of the next bond to break (noted $\AAA$, see Fig. \ref{fig:Sketch}). As before, the crack tip is first placed, as a first guess at the middle $\OO$ of the next bond about to break. As $L \rightarrow \infty$, the only terms that survive in Williams' expansion for displacement fields (Eq. \ref{eqWilliams}) are the $n=-1$ and $n=1$ terms, where the former is attributed to the crack tip mispositioning: 

\begin{equation}
u=a_{-1}\Phi_{-1}^{III}(r,\theta) + a_{1}\Phi_{1}^{III}(r,\theta),
\label{S6:equ1}
\end{equation}

\noindent Kirchhoff's law then imposes $\sum_i \left(u(\AAA)-u(\AAA_i)\right)=0$, where $\AAA_i$ are the nodes directly connected to $\AAA$ (Fig. \ref{fig:Sketch}). This leads:

\begin{equation}
a_{-1} S_{-1}(\AAA,\AAA_i) + a_{1} S_{1}(\AAA,\AAA_i)=0,
\label{S6:equ2}
\end{equation}

\noindent where:

\begin{equation}
S_n(\AAA,\AAA_i)=\sum_i\left(\Phi_n^{III}(r_\AAA,\theta_\AAA)-\Phi_n^{III}(r_{\AAA_i},\theta_{\AAA_i})\right).
\label{S6:equ3}
\end{equation}

As in Sec. 3.2, a first-order approximation of the mispositioning is then given by $d=2 a_{-1}/a_1$:

\begin{equation}
d= -2\frac{S_{1}(\AAA,\AAA_i)}{S_{-1}(\AAA,\AAA_i)}.
\label{S6:equ4}
\end{equation}

\noindent The frame origin $O$ is then shifted to the new position $O \rightarrow O - d.\vec{e}_x$. Then, the displacement fields $u(\AAA)$ and $u(\AAA_i)$ are computed again, the mispositioning is computed again and the frame origin is switched again. Figure \ref{fig:d_Vs_Number_Itterration} shows the variation of this mispositioning as a function of the number of iterations.  After $6$ iterations, the mispositioning converges toward a constant value, $d_{theo}(1^{st}order)$, that depends only on the lattice geometry and system size. The values obtained for the three lattice geometries studied here are reported in the second line of Tab. \ref{tab_d}.

Once the mispositioning $d_{theo}(1^{st}order)$ is known and the crack tip is placed in its proper location, $\CC$ (Fig. \ref{fig:Sketch}), the $n=-1$ super-singular term vanishes and the displacement field writes:

\begin{equation}
u=a_{1}\Phi_{1}^{III}(\tilde{r},\tilde{\theta}),
\label{S6:equ5}
\end{equation}

\noindent where $(\tilde{r},\tilde{\theta})$ refers to the polar coordinates in the frame now centered at $\CC$. $a_1$ relates to $K$ via Eq. \ref{eqK} and the fracture toughness $K_c$ can be determined by stating that $K=K_c$ when the force $i_{\AAA \BB}=u(\AAA)-u(\BB)$ applying to the next bond to beak (bond $\AAA-\BB$ in Fig. \ref{fig:Sketch}) is equal to the breaking threshold $i_c=\sqrt{2}$. This yields: 

\begin{equation}
K_c=\frac{E \sqrt{\pi}}{2\left(\Phi^{III}_1(\tilde{r}_\AAA,\tilde{\theta}_\AAA)-\Phi^{III}_1(\tilde{r}_\BB,\tilde{\theta}_\BB)\right)}
\label{S6:equ6}
\end{equation}

\noindent Fracture energy, $\Gamma_{theo}(1^{st}order)$, is subsequently deduced by applying Irwin's formula (Eq. \ref{eqIrwin}). The values obtained for the three lattice geometries studied here are reported in the fourth line of Tab. \ref{tab_d}.

Figure \ref{fig:13} shows the relative differences between $d(L)$ (Top panel) and $\Gamma(L)$ (Bottom panel) as measured in the simulations, and the analytical values $d_{theo}(1^{st}order)$ and $\Gamma_{theo}(1^{st}order)$ as determined via the 1$^{st}$ order theory presented here. In both cases, the differences are plotted as a function of the lattice size $L$. In all cases, the differences decrease as system size increases, as expected since theory assumes $L \rightarrow \infty$ so that all terms $n > 1$ could be neglected in Williams' expansion (Eq. \ref{eqWilliams}). Conversely, the differences do not vanish as $L \rightarrow \infty$: In particular, the relative difference between $\Gamma^\infty$ and $\Gamma_{theo}(1^{st}order)$  is about $16\%$, $4.7\%$ and $12.8\%$ for honeycomb, triangular and square lattices, respectively. Something is still missing in the theory. 

\subsection{Effect of higher order super-singular terms}
\label{sec:AnalyticalMethod}

\begin{figure}
\includegraphics[width=0.5\textwidth]{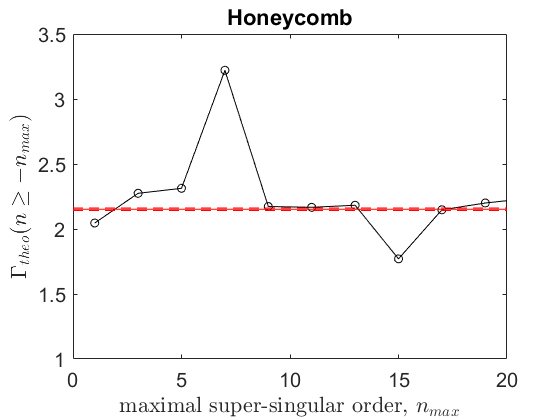}
\hfill
\includegraphics[width=0.5\textwidth]{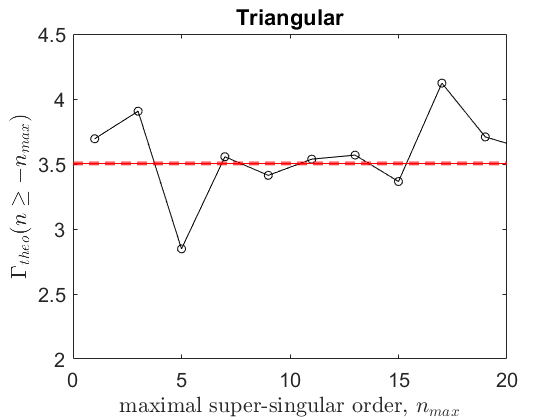}
\hfill
\includegraphics[width=0.5\textwidth]{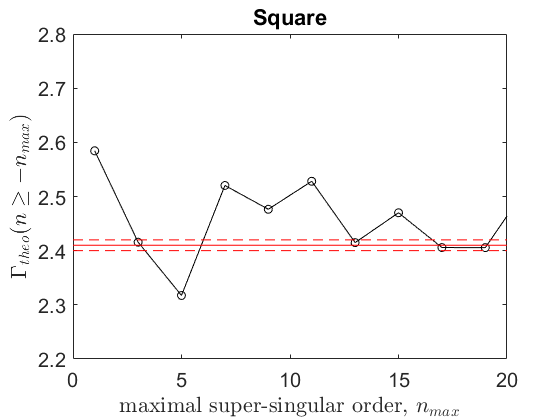}
\caption{Fracture toughness calculated by the analytical method proposed in Sec. 6.2 by including the higher order super-singular terms $n \geq n_{max}$. The red lines show the fracture toughness values calculated by simulations. $\Gamma_{theo}$ is expressed in $\gamma/\ell$ unit.}
\label{fig:Gamma_theo_exact}
\end{figure}

The above analysis assumes that all super-singular ($n < 0$) terms vanish in Williams' expansion once the crack tip is properly located. However, these terms may be relevant here since we are considering discrete lattices. Moreover, if these terms were to exist, they are dominant since we are examining the near-tip displacement field, in the limit where $r$ is very small. 

We propose then to place arbitrarily the crack tip at the middle $\OO$ of the next bond to break, as in section 6.1, and to consider the other super-singular terms in Williams' expansion. Equation \ref{S6:equ1} becomes: 

\begin{equation}
u=\sum_{k=1}^{k_{max}} a_{1-2k}\Phi_{1-2k}^{III}(r,\theta) + a_{1}\Phi_{1}^{III}(r,\theta),
\label{S6:equ7}
\end{equation}

\noindent where $n_{max}=2k_{max}-1$ defines the maximal super-singular order. As before, it is the application of Kirchhoff's laws at some nodes $\PPP$ that allows the determination of the super-singular terms:

\begin{equation}
\sum_{k=1}^{k_{max}} d_{1-2k} S_{1-2k}(\PPP,\PPP_i) = - 2 S_1(\PPP,\PPP_i), 
\label{S3:equ8}
\end{equation}

\noindent where $d_n = a_n/a_1$ is the relative contribution of the super-singular term $n$ with respect to the singular term $a_1$. Here and as in previous section, $S_{n}(\PPP,\PPP_i)$ is given by Eq. \ref{S6:equ3}. To find the $n_{max}$ unknowns, $d_n$, we applied the above Kirchhoff's law to the $n_{max}$ nodes $\PPP$ the closest to the crack-tip. This provides a well-posed system of linear equations. Once it is solved and the $d_n$ are determined, as before, the fracture toughness can be determined by stating that $K=K_c$ when the force $i_{\AAA\BB}=u(\BB)-u(\AAA)$ applying to the next bond to break is equal to the breaking threshold. This yields: 

\begin{strip}
\begin{equation}
K_c=\frac{E \sqrt{\pi}}{\sum_{k=1}^{k_{max}} d_{1-2k}\left(\Phi^{III}_{1-2k}(r_\AAA,\theta_\AAA)-\Phi^{III}_{1-2k}(r_\BB,\theta_\BB)\right)+2\left(\Phi^{III}_1(r_\AAA,\theta_\AAA)-\Phi^{III}_1(r_\BB,\theta_\BB)\right)}
\label{S6:equ9}
\end{equation}
\end{strip}

\noindent Fracture energy, $\Gamma_{theo}(n\geq n_{max})$, is subsequently deduced by applying Irwin's formula (Eq. \ref{eqIrwin}).

Figure \ref{fig:Gamma_theo_exact} shows the evolution of fracture energy with $n_{max}$ for the three lattice geometry examined. Significant variations are observed. In all cases, the horizontal red line indicates the value obtained numerically.  

\section{Concluding discussion}
\label{sec:concl}

This numerical study was designed to unravel how lattice geometry drives fracture toughness in 2D fuse networks, which are analogs of 2D brittle crystals under antiplanar loading. The main observations are:
\begin{itemize}
    \item Irwin's relation between fracture energy and fracture toughness is fulfilled as soon as the system size is large enough, that is $L \geq 50$;
    \item Significant size dependencies are observed. Both fracture toughness and fracture energy converge algebraically (associated exponent close to unity) toward loading-independent material-constant values as system size tends to infinity; 
    \item The convergence speed is faster when the lattice geometry increases;
    \item The material-constant fracture energy determined in the simulations in the limit of infinite lattices is significantly larger than Griffith's free surface energy, $2 \gamma_S$;
    \item Conversely, the material-constant fracture toughness determined in the simulations in the limit of infinite lattices can be approached up to $\sim 15\%$ using the analytical procedure proposed by \cite{Nguyen19_prl};
    \item The residual errors between the numerical determination and the analytical procedure above are a consequence of lattice discreetness (which is not considered in LEFM) and can be decreased by introducing an arbitrary number of super-singular terms in Williams' solutions.
\end{itemize}

This work emphasizes the role played by lattice discreetness in the selection of the macroscale resistance to fracture. Note that the analysis presented here concerns fracture {\em initiation} toughness, $K_{ci}$, that is the value of the stress intensity factor at the moment when the force acting on the next element to break is equal to the breakdown value. Similarly, the fracture energy discussed throughout this work is the fracture energy at initiation, $\Gamma_{i}$. Once crack growth commences, inertia effects come into play, elastic waves are emitted and a full dynamic analysis is required to determine fracture {\em growth} energy, $\Gamma_g$. This analysis has been performed by \cite{Slepyan81_spd} for semi-infinite crack growing at a constant speed, $v$, in a 2D square lattice under antiplanar loading, and by \cite{Kulakhmetova84_ms} for a tensile crack moving in a triangular lattice. They characterized the structure and energy of the waves emitted during the fracture and determined the evolution of $\Gamma_g(v)$ in both cases. They showed that $\Gamma_g(v)$ is larger than Griffith surface energy $2 \gamma_S$ over the whole velocity range including the $v \rightarrow 0$ limit; the excess of $\Gamma_g(v \rightarrow 0)$ over $2\gamma_S$ reflects the existence of phonons \citep{Slepyan10_ap}.

It is of interest to compare fracture initiation energy $\Gamma_i$ with fracture growth energy at vanishing speed $\Gamma_g(v \rightarrow 0)$. {\em A priori}, they are not the same; continuum fracture mechanics commonly states that $\Gamma_i > \Gamma_g(v \rightarrow 0)$, so that crack jumps to a finite speed immediately upon initiation \citep{Ravichandar04_book}. Remarkably, this is not the case in the mode III 2D fracture lattice problem: The value $\Gamma_i = 2.409 \pm 0.009$ we determined here via static analysis in a square lattice is nearly the same as that obtained by \cite{Slepyan81_spd} for $\Gamma_g(v \rightarrow 0)$ via dynamic analysis in the same lattice geometry. This suggests that, once a crack starts propagating in the lattice, the phonon structure can accommodate the fracture energy in excess with respect to surface energy, without necessarily requiring a finite fracture speed. 

The analytical procedure proposed by \cite{Nguyen19_prl} has been shown here to provide fairly accurate fracture toughness predictions for antiplanar lattice problems. Its extension to beam lattices is quite straightforward and, as such, it could provide a promising tool to determine the fracture toughness in micro-/nano-lattices \citep{Schaedler11_science, Zheng14_science, Zhang19_small} formed by periodically arranged sets of microscale/nanoscale beams or tubes. This novel class of metamaterials, indeed, has revealed outstanding mechanical performances, with ultralow densities, high stiffness, and large recoverability upon compression, but unusual aspects in the way to assess their fracture toughness \citep{Shaikeea22_natmat}. Work in this direction is in progress.

\renewcommand{\thefigure}{A\arabic{figure}}
\renewcommand{\theequation}{A\arabic{equation}}
\setcounter{figure}{0}
\setcounter{equation}{0}

\appendix
\section{Stress field analysis}\label{A1}

\begin{figure*}[htp]
\includegraphics[width=0.99\textwidth]{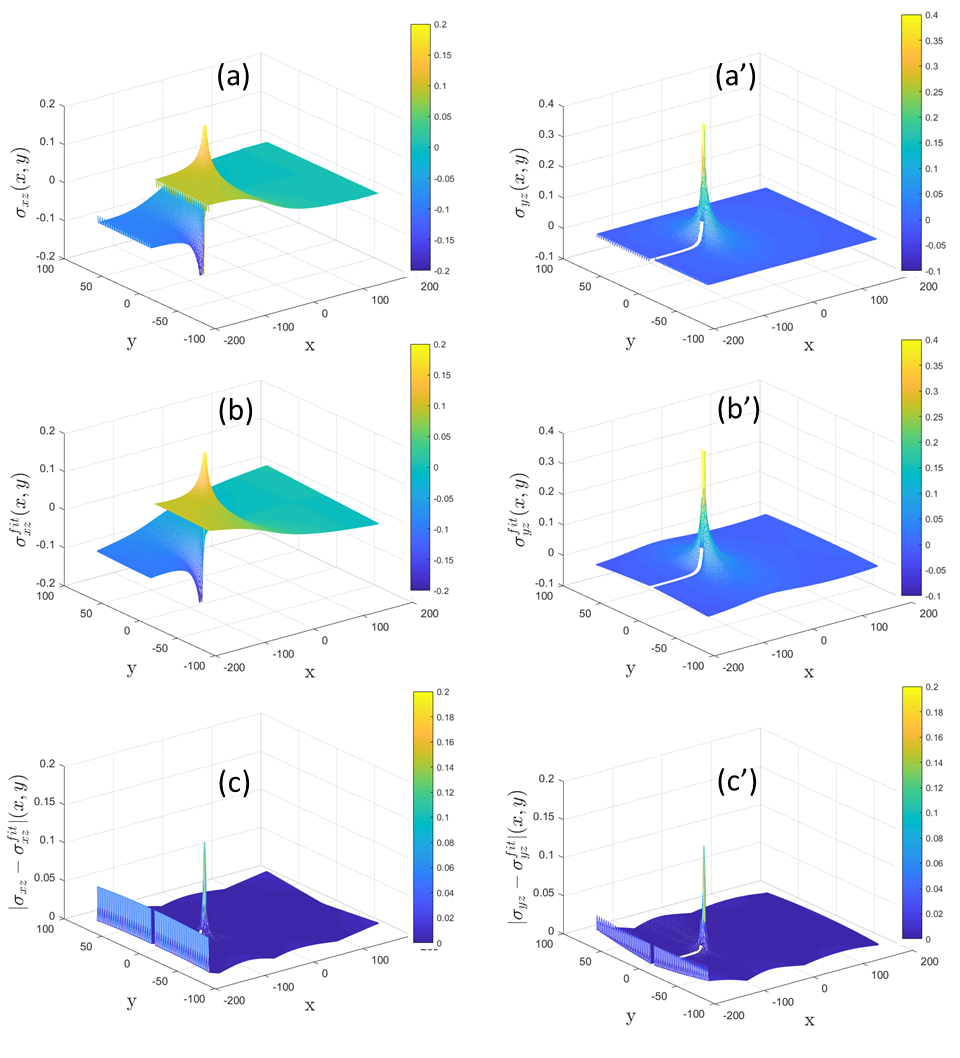}
\caption{
Comparison between virial stress and Williams' series for stress field: Panels {\bf a} and {\bf a'} respectively show the two non-null components of the stress field, $\sigma_{xz}(x,y)$ and $\sigma_{yz}(x,y)$, observed the simulation of a $400 \times 200$ honeycomb lattice under CT loading. The stress field has been computed via Eq. \ref{A:equ1}. Panels {\bf b} and {\bf b'} show the predicted fields, $\sigma^{fit}_{xz}(x,y)$ and $\sigma^{fit}_{yz}(x,y)$ using Eqs. \ref{eqWilliamsStress} truncated to and with $n \leq 9$) with a tip origin (crack tip position) located to the position $\CC$ determined by applying the procedure described in section 3.2 onto the displacement field in the same lattice, and coefficients $\{a_n\}$ obtained via fitting this displacement field. 
Panels {\bf c} and {\bf c'} show the absolute difference, $|\sigma^{fit}_{xz}-\sigma_{xz}|(x,y)$ and $|\sigma^{fit}_{yz}-\sigma_{xz}|(x,y)$. The fit is very good everywhere, except in the very vicinity of the crack tip.  All quantities in the panels are expressed in reduced units: $\ell$ unit for positions $x$ and $y$,  $g$ unit for stresses $\sigma_{xz}(x, y)$, $\sigma_{yz}(x, y)$, $\sigma^{fit}_{xz}(x, y)$, $\sigma^{fit}_{yz}(x, y)$.
}
\label{fig:S1}
\end{figure*}

Voltage field, $u(x,y)$, in the 2D fuse networks investigated here can be identified with the out-of-plane displacement $\uu = u \ee_z $ in 2D lattices of the same geometry under antiplanar loading. Accordingly, the current $i_{\PPP \PPP'} = u(\PPP) - u(\PPP')$ flowing from $\PPP'$ to $\PPP$ through the bond $\PPP - \PPP'$ can be identified with the force exerted by this bond on node $\PPP$. Virial stress can then be employed to compute the local mechanical stress tensor applying on the Voronoi cell surrounding this node $\PPP$, ${\bf \sigma}(\PPP)$. The two non-zero components are then given by: 

\begin{align}
\begin{split}
\sigma_{xz}(\PPP) = \frac{1}{2 Vor}\sum_i (u(\PPP) - u(\PPP_i)) (x_{\PPP} - x_{\PPP_i}),\\
\sigma_{yz}(\PPP) = \frac{1}{2 Vor}\sum_i (u(\PPP) - u(\PPP_i)) (y_{\PPP} - y_{\PPP_i}).
\label{A:equ1}
\end{split}
\end{align}

\noindent Here, $\PPP_i$ refer to the nodes connected to $\PPP$, and $Vor$ is the area of the Voronoi polyhedra associated to $\PPP$: $Vor=1$, $\sqrt{3}/2$, and $3\sqrt{3}/4$ in the square, triangular, and honeycomb lattices, respectively. We plotted, in Figs. \ref{fig:S1}(a) and \ref{fig:S1}(a'), these two stress field components for the same $400 \times 200$ noneycomb lattice as that of Fig. \ref{fig:2}. Note, in particular, the divergence of the stress field at the crack tip. Actually, Williams's Eqs. \ref{eqWilliams} and \ref{eqWilliamsScal} can be employed to express stress field as a series of elementary functions. Indeed, $\sigma_{xz}$ and $\sigma_{yz}$ write:

\begin{equation}
\sigma_{xz} = E \epsilon_{xz} = \frac{E}{2} \frac{\partial u}{\partial x}, \quad
\sigma_{yz} = E \epsilon_{yz} = \frac{E}{2} \frac{\partial u}{\partial y}.
\end{equation}
  
\noindent By introducing Eqs. \ref{eqWilliams} and \ref{eqWilliamsScal} in the above equation, and then by noting that $\frac{\partial}{\partial x} = \cos\theta\frac{\partial}{\partial r} -\frac{\sin\theta}{r} \frac{\partial}{\partial \theta}$ and $\frac{\partial}{\partial y} = \sin\theta \frac{\partial}{\partial r} + \frac{\cos\theta}{r}\frac{\partial}{\partial \theta}$, one gets:  

\begin{align}
\begin{split}
\sigma_{xz} = \frac{E}{2} \sum_{n\geq 0}  a_n \frac{n}{2} r^{\frac{n}{2}-1}\sin\left(\frac{n}{2}-1\right)\theta,\\
\sigma_{yz} = \frac{E}{2} \sum_{n\geq 0}  a_n \frac{n}{2} r^{\frac{n}{2}-1}\cos \left(\frac{n}{2}-1\right)\theta.
\end{split}
\label{eqWilliamsStress}
\end{align}

Figures \ref{fig:S1}(b) and \ref{fig:S1}(b') show the Williams' stress field as obtained using the equation above with the tip position, $\CC$, and parameters, $\{a_n\}$, determined from the fit of Williams' displacement field in Fig. \ref{fig:2}.
 Note the apparent agreement with the fields observed on the numerical simulations (Figs. \ref{fig:S1}(a) and \ref{fig:S1}(a')). Figures \ref{fig:S1}(c) and \ref{fig:S1}(c')  show the absolute difference between numerical and Williams' fields. The fit is very good, except in the very vicinity of the crack tip.

\begin{acknowledgements}
\end{acknowledgements}

\section*{Contributions}
T.N. and D.B. contributed equally to this work

\section*{Conflict of interest}
The authors declare that they have no conflict of interest.



%\bibliography{biblio}

\begin{thebibliography}{29}
\providecommand{\natexlab}[1]{#1}
\providecommand{\url}[1]{{#1}}
\providecommand{\urlprefix}{URL }
\expandafter\ifx\csname urlstyle\endcsname\relax
  \providecommand{\doi}[1]{DOI~\discretionary{}{}{}#1}\else
  \providecommand{\doi}{DOI~\discretionary{}{}{}\begingroup
  \urlstyle{rm}\Url}\fi
\providecommand{\eprint}[2][]{\url{#2}}

\bibitem[{Bernstein and Hess(2003)}]{Bernstein03_prl}
Bernstein N, Hess D (2003) Lattice trapping barriers to brittle fracture.
  Physical review letters 91(2):025501

\bibitem[{Bonamy(2017)}]{Bonamy17_crp}
Bonamy D (2017) Dynamics of cracks in disordered materials. Comptes Rendus
  Physique 18(5-6): 297--313

\bibitem[{Buehler et~al.(2007)Buehler, Tang, van Duin, and
  Goddard}]{Buehler07_prl}
Buehler MJ, Tang H, van Duin AC, Goddard WA (2007) Threshold crack speed
  controls dynamical fracture of silicon single crystals. Physical Review
  Letters 99(16):165502

\bibitem[{De-Arcangelis and Herrmann(1989)}]{Arcangelis89}
De-Arcangelis L, Herrmann HJ (1989) Scaling and multiscaling laws in random
  fuse networks. Physical Review B 39:2678--2684

\bibitem[{Gordon(1978)}]{Gordon78}
Gordon J (1978) Structures, or Why Things Don't Fall Down. Plenum, New York

\bibitem[{Griffith(1921)}]{Griffith20_ptrs}
Griffith AA (1921) The phenomena of rupture and flow in solids. Philosophical
  Transaction of the Royal Society of London A221:163

\bibitem[{Hansen et~al.(1991)Hansen, Hinrichsen, and Roux}]{Hansen91_prl}
Hansen A, Hinrichsen H, Roux S (1991) Roughness of crack interfaces. Physical
  Review Letters 66(19):2476--2479

\bibitem[{Hsieh and Thomson(1973)}]{Hsieh73_jap}
Hsieh C, Thomson R (1973) Lattice theory of fracture and crack creep. Journal
  of Applied Physics 44(5):2051--2063

\bibitem[{Irwin(1957)}]{Irwin57_jam}
Irwin GR (1957) Analysis of stresses and strains near the end of a crack
  traversing a plate. Journal of Applied Mechanics 24:361

\bibitem[{Kermode et~al.(2015)Kermode, Gleizer, Kovel, Pastewka, Cs{\'a}nyi,
  Sherman, and De~Vita}]{Kermode15_prl}
Kermode JR, Gleizer A, Kovel G, Pastewka L, Cs{\'a}nyi G, Sherman D, De~Vita A
  (2015) Low speed crack propagation via kink formation and advance on the
  silicon (110) cleavage plane. Physical review letters 115(13):135501

\bibitem[{Kulakhmetova et~al.(1984)Kulakhmetova, Saraikin, and
  Slepyan}]{Kulakhmetova84_ms}
Kulakhmetova SA, Saraikin VA, Slepyan LI (1984) Plane problem of a crack in a
  lattice. Mechanics of Solids 19:101

\bibitem[{Lawn(1993)}]{Lawn93_book}
Lawn B (1993) fracture of brittle solids. Cambridge solide state science

\bibitem[{Marder and Gross(1995)}]{Marder95_jmps}
Marder M, Gross S (1995) Origin of crack tip instabilities. Journal of the
  Mechanics and Physics of Solids 43(1):1--48,
  \doi{https://doi.org/10.1016/0022-5096(94)00060-I},
  \urlprefix\url{https://www.sciencedirect.com/science/article/pii/002250969400060I}

\bibitem[{Nguyen and Bonamy(2019)}]{Nguyen19_prl}
Nguyen T, Bonamy D (2019) Role of the crystal lattice structure in predicting
  fracture toughness. Physical review letters 123:205503

\bibitem[{P{\'e}rez and Gumbsch(2000)}]{Perez00_prl}
P{\'e}rez R, Gumbsch P (2000) Directional anisotropy in the cleavage fracture
  of silicon. Physical review letters 84(23):5347

\bibitem[{Ravi-Chandar(2004)}]{Ravichandar04_book}
Ravi-Chandar K (2004) Dynamic Fracture. Elsevier Ltd

\bibitem[{R{\'e}thor{\'e} and Estevez(2013)}]{Rethore13_jmps}
R{\'e}thor{\'e} J, Estevez R (2013) Identification of a cohesive zone model
  from digital images at the micron-scale. Journal of the Mechanics and Physics
  of Solids 61(6):1407--1420

\bibitem[{Santucci et~al.(2004)Santucci, Vanel, and Ciliberto}]{Santucci04_prl}
Santucci S, Vanel L, Ciliberto S (2004) Subcritical statistics in rupture of
  fibrous materials: Experiments and model. Physical Review Letters 93:095505

\bibitem[{Schaedler et~al.(2011)Schaedler, Jacobsen, Torrents, Sorensen, Lian,
  Greer, Valdevit, and Carter}]{Schaedler11_science}
Schaedler TA, Jacobsen AJ, Torrents A, Sorensen AE, Lian J, Greer JR, Valdevit
  L, Carter WB (2011) Ultralight metallic microlattices. Science
  334(6058):962--965, \doi{10.1126/science.1211649}

\bibitem[{Shaikeea et~al.(2022)Shaikeea, Cui, O'Masta, Zheng, and
  Deshpande}]{Shaikeea22_natmat}
Shaikeea AJD, Cui H, O'Masta M, Zheng XR, Deshpande VS (2022) The toughness of
  mechanical metamaterials. Nature Materials 21(3):297--304,
  \doi{10.1038/s41563-021-01182-1}

\bibitem[{Slepyan(1981)}]{Slepyan81_spd}
Slepyan LI (1981) Dynamics of a crack in a lattice. Sov Phys Dokl 26:538

\bibitem[{Slepyan(2010)}]{Slepyan10_ap}
Slepyan LI (2010) Wave radiation in lattice fracture. Acoustical Physics
  56(6):962--971, \doi{10.1134/s1063771010060217}

\bibitem[{Thomson et~al.(1971)Thomson, Hsieh, and Rana}]{Thomson71_jap}
Thomson R, Hsieh C, Rana V (1971) Lattice trapping of fracture cracks. Journal
  of Applied Physics 42(8):3154--3160

\bibitem[{Williams(1952)}]{Williams52_jam}
Williams ML (1952) Journal of Applied Mechanics 19:526--528

\bibitem[{Zapperi et~al.(1997)Zapperi, Vespignani, and
  Stanley}]{Zapperi97_nature}
Zapperi S, Vespignani A, Stanley HE (1997) Plasticity and avalanche behaviour
  in microfracturing phenomena. Nature 388:658--660,
  \urlprefix\url{http://www.nature.com/nature/journal/v388/n6643/full/388658a0.html}

\bibitem[{Zapperi et~al.(2005)Zapperi, Nukala, and Simunovic}]{Zapperi05_pre}
Zapperi S, Nukala PKVV, Simunovic S (2005) Crack roughness and avalanche
  precursors in the random fuse model. Physical Review E 71(2 Pt 2):026106

\bibitem[{Zhang et~al.(2019)Zhang, Wang, Ding, and Li}]{Zhang19_small}
Zhang X, Wang Y, Ding B, Li X (2019) Design, fabrication, and mechanics of 3d
  micro-/nanolattices. Small 16(15):1902842, \doi{10.1002/smll.201902842}

\bibitem[{Zheng et~al.(2014)Zheng, Lee, Weisgraber, Shusteff, DeOtte, Duoss,
  Kuntz, Biener, Ge, Jackson, Kucheyev, Fang, and Spadaccini}]{Zheng14_science}
Zheng X, Lee H, Weisgraber TH, Shusteff M, DeOtte J, Duoss EB, Kuntz JD, Biener
  MM, Ge Q, Jackson JA, Kucheyev SO, Fang NX, Spadaccini CM (2014) Ultralight,
  ultrastiff mechanical metamaterials. Science 344(6190):1373--1377,
  \doi{10.1126/science.1252291}

\bibitem[{Zhu et~al.(2004)Zhu, Li, and Yip}]{Zhu04_prl}
Zhu T, Li J, Yip S (2004) Atomistic study of dislocation loop emission from a
  crack tip. Physical review letters 93(2):025503

\end{thebibliography}
\end{document}